\documentclass[10pt, pre, amsmath, onecolumn, showpacs, superscriptaddress, nofootinbib]{revtex4-1}

\usepackage{amsmath, amsthm, amssymb, amsfonts}
\usepackage{mathtools}
\usepackage{hyperref}
\hypersetup{
    colorlinks=true,       
    linkcolor=red,          
    citecolor=blue,        
    filecolor=magenta,      
    urlcolor=cyan           
}
\usepackage{graphicx}
\usepackage{dsfont}
\usepackage{fullpage}
\usepackage{color}
\usepackage{soul} 
\usepackage[title]{appendix}
\usepackage{tikz}
\usetikzlibrary{patterns}
\usetikzlibrary{shapes.multipart}
\usetikzlibrary{arrows}

\theoremstyle{definition}

\definecolor{labelkey}{cmyk}{.4,.2,0,0}

\newcommand{\red}{\color{red}}
\newcommand{\blue}{\normalcolor}
\newcommand{\be}{\begin{equation}}
\newcommand{\ee}{\end{equation}}
\newcommand{\beq}{\begin{equation}}
\newcommand{\eeq}{\end{equation}}
\newcommand{\bea}{\begin{eqnarray}}
\newcommand{\eea}{\end{eqnarray}}
\newcommand{\nn}{\nonumber}

\begin{document}

\title{\bf \large Non-uniqueness of the steady state for run-and-tumble particles with a double-well interaction potential}
\date{\today}

\author{L\'eo \surname{Touzo}}
\affiliation{Laboratoire de Physique de l'Ecole Normale Sup\'erieure, CNRS, ENS and PSL Universit\'e, Sorbonne Universit\'e, Universit\'e Paris Cit\'e, 24 rue Lhomond, 75005 Paris, France}
\affiliation{University of Chicago, James Franck Institute, 929 E 57th Street, Chicago, IL 60637
}
\author{Pierre Le Doussal}
\affiliation{Laboratoire de Physique de l'Ecole Normale Sup\'erieure, CNRS, ENS and PSL Universit\'e, Sorbonne Universit\'e, Universit\'e Paris Cit\'e, 24 rue Lhomond, 75005 Paris, France}

\date{\today}

\begin{abstract}
We study $N$ run-and-tumble particles (RTPs) in one dimension interacting via a double-well pairwise potential $W(r)=-k_0 \, r^2/2+g \, r^4/4$, which is repulsive at short interparticle distance $r$ and attractive at large distance. At large time, the system forms a bound state where the density of particles has a finite support. We focus on the determination of the total density of particles in the stationary state $\rho_s(x)$, in the limit $N\to+\infty$. We obtain an explicit expression for $\rho_s(x)$ as a function of the ``renormalized" interaction parameter $k=k_0-3m_2$ where $m_2$ is the second moment of $\rho_s(x)$. Interestingly, this stationary solution exhibits a transition between a connected and a disconnected support for a certain value of $k$, which has no equivalent in the case of Brownian particles. Analyzing in detail the expression of the stationary density in the two cases, we find a variety of regimes characterized by different behaviors near the edges of the support and around $x=0$. Furthermore, by studying the relation between $k$ and $k_0$, we find that the mapping $k_0\to k$ becomes multi-valued below a certain value of the tumbling rate $\gamma$ of the RTPs for some range of values of $k_0$ near the transition, implying the existence of two stable solutions. Finally, we show that in the case of a disconnected support, it is possible to observe steady states where the density $\rho_s(x)$ is not symmetric, characterized by a third moment $m_3$ which can take a continuous range of values. All our analytical predictions are in good agreement with numerical simulations already for systems of $N = 100$ particles. The non-uniqueness of the stationary state is a particular feature of this model in the presence of active (RTP) noise, which contrasts with the uniqueness of the Gibbs equilibrium for Brownian particles. We argue that these results are also relevant for a class of more realistic interactions with both an attractive and a repulsive part, but which decay at infinity.
\end{abstract}

\maketitle


\section{Introduction and main results}

\subsection{Overview}

There is much current interest in the study of interacting active particles \cite{soft,Ramaswamy2017,BechingerRev,Marchetti2018,Cates2012}. Due to their intrinsically out-of-equilibrium nature, such systems exhibit new types of phase transitions, such as motility-induced phase separation, i.e. a separation between a dense and a dilute phase in the presence of short-range repulsive interactions \cite{FM2012,Buttinoni2013,FHM2014,CT2015,OByrne2021,Active_OU}. To study such systems analytically, hydrodynamic approaches and perturbative methods have been developed \cite{TailleurCates,Active_OU,Bialke2013,Wittkowski2014,Solon2018_1,Solon2018_2}. However, there are currently very few exact results available, even for one dimensional systems. Notable exceptions include the two-particle case \cite{slowman,slowman2,us_bound_state,Maes_bound_state,MBE2019,KunduGap2020,Metson2022,MetsonLong,Hahn2023,Hahn2025}, 
harmonic chains \cite{SinghChain2020,PutBerxVanderzande2019, HarmonicChainRevABP,HarmonicChainRTPDhar}, as well as some specific lattice models with contact interactions~\cite{KH2018,Agranov2021,Agranov2022,lattice2lanes2025,Dandekar2020,Thom2011,nonexistence}.

In the continuum, one of the simplest models of active particle is the run-and-tumble particle (RTP), inspired by the motion of E. Coli bacteria \cite{Berg2004}. In one dimension, it is
driven by so-called telegraphic noise, which alternates between two values at a constant rate \cite{HJ95,W02,ML17,kac74,Orshinger90,Schnitzer,TailleurCates}.
Recently we have considered a system of $N$ RTPs in one dimension, interacting
via a pairwise power law potential \cite{RieszFluct}.
In particular we have studied two cases of long range interactions,
the logarithmic potential, which can be seen as an active generalization
of the Dyson Brownian motion \cite{TouzoDBM2023}, and the linear 1D Coulomb potential, also
called active rank diffusions \cite{activeRDshort,activeRDlong}. In both cases, the aim was to compute the 
density of particles in the steady state in the limit of large $N$. In the case of an attractive 1D Coulomb interaction, one finds that the particles form a bound state which exhibits a transition between a smooth density with unbounded support, and a density with bounded support displaying clusters at the edges
\cite{activeRDshort,activeRDlong}.

Another form of potential interaction which is often considered involves a repulsive part at short interparticle distance and an attractive part at large distance, such as the Lennard-Jones potential
which decreases to zero at infinity. For active particles in two space dimensions, such interactions have been shown to lead to a reentrant phase transition \cite{Poon2012,Redner2013}.

{\blue In this paper we consider a toy model for such interactions, which consists in $N$ one-dimensional RTPs interacting via a pairwise polynomial potential of the form $W(r)=-k_0 \, r^2/2+g \, r^4/4$, which for $k_0,g>0$ is repulsive at short distance
and attractive at large distance. We find that in the steady state the system forms a bound state with a bounded support, which
exhibits a transition between a joint support and a disconnected support where particles
spontaneously split into two groups. Furthermore, remarkably, we find that the 
steady state is not unique, and that in some range of parameters it exhibits 
bistability, as well as a breaking of the parity symmetry. The steady state which 
is reached at large time is found to depend on some features
of the initial conditions. By contrast, both of these features are absent in the case of Brownian particles, for which the Gibbs equilibrium is unique. We note that, although this work focuses on the large $N$ limit of the model, the existence of multiple steady states and the breaking of parity symmetry are already visible for small values of $N$, as we have observed from numerical simulations, which we report in Appendix \ref{app:N2}.

Our model has two advantages from a computational perspective.
First, $W'(0)=0$, hence it allows for particle crossings and therefore allows to use the extension of the
Dean-Kawasaki method \cite{Dean,Kawa} to the RTP (while it was shown that this method fails to describe
single-file active particles \cite{TouzoDBM2023}). Second, it is simple enough so
that the self-consistent equation obeyed by the stationary density in the large $N$ limit, derived in \cite{activeRDshort},
can be solved analytically. Furthermore, although our model has an interaction force $-W'(r)$ which diverges at large distance, which is a priori an unrealistic feature, one can make an argument that a similar behavior can be observed 
for more realistic interaction forces which vanish at infinity, but for which the system still has a bound state. A more precise criterion for the existence of a bound state and for
the transition from single to disjoint support 
is discussed in Appendix \ref{app:realistic}, and is confirmed by a 
numerical simulation. In that case we expect all the results obtained for the double well toy model
to be qualitatively relevant. }

We now describe the model in more details, present the general method for the solution, 
and summarize the main results.

{\red 







}

\subsection{Model and methodology}

We consider $N$ run-and-tumble particles in one dimension 
interacting via a two-body potential $W(x)$, which can be described by the equations of motion
\be \label{langevin_doublewell}
\frac{dx_i}{dt} = -\frac{1}{N} \sum_{j=1}^N W'(x_i-x_j) + v_0 \sigma_i(t) + \sqrt{2 T} \xi_i(t) \;,
\ee 
where the $\sigma_i(t)$ are i.i.d. telegraphic noises, which switch between values $\pm 1$ with a rate $\gamma$ (called the tumbling rate) and the $\xi_i(t)$ are i.i.d. Gaussian white noises with unit variance. 
In the case where the interaction potential $W(x)$ is sufficiently attractive, 
the stationary state has the form of a bound state. One defines 
the time-dependent densities, with $\sigma=\pm 1$
\be
\rho_\sigma(x,t) = \frac{1}{N} \sum_i \delta(x-x_i(t)) \delta_{\sigma,\sigma_i(t)} \;,
\ee
as well as the total density $\rho_s$ and the ``magnetization" $\rho_d$,
\be 
\rho_s(x,t) = \rho_+(x,t) + \rho_-(x,t) \quad , \quad \rho_d(x,t) = \rho_+(x,t) - \rho_-(x,t) \;.
\ee 
If $W'(0)=0$, the particles can always cross even for $T=0$, and one can use the Dean-Kawasaki method \cite{Dean,Kawa} for 1D RTPs introduced in \cite{TouzoDBM2023}. In the large $N$ limit, for $T=0$, we have shown in \cite{activeRDshort} (the derivation is recalled in Appendix~\ref{app:self}) that the stationary density obeys the self-consistent equation
\be 
\rho_s(x) = \frac{K}{v_0^2-\tilde F(x)^2} e^{2 \gamma \int^x dz \frac{\tilde F(z)}{v_0^2-\tilde F(z)^2}}
\quad , \quad 
\tilde F(z)= - \int dy \, W'(z-y) \rho_s(y) \;,
\label{self_intro}
\ee 
where $K$ is a constant determined by normalization. Here $\tilde F(x)$ is an effective force
field, which depends itself on the density. Once the total density $\rho_s(x)$ is known, the 
density $\rho_d(x)$ is obtained as 
\be \label{rhod} 
\rho_d(x) = -\frac{\tilde F(x)}{v_0} \rho_s(x) \;.
\ee
The equation \eqref{self_intro} holds when the support of $\rho_s(x)$ is a single interval,
but can be generalized to the case where the support is a union of disjoint intervals.
This is precisely the situation that we will study below.

In the present paper, we will focus on an example of interaction $W(x)$ where
these self-consistent equations can be solved explicitly, namely the
case of a double well potential 
\be \label{doublewell}
W(x) = -  k_0 \frac{x^2}{2} + g \frac{x^4}{4} 
\ee 
(which indeed satisfies $W'(0)=0$). We focus on the case $g>0$, for which the interaction is attractive at large distance.
The double well potential is of particular interest since it can exhibit a phase transition
where the bound state splits into two spatially disconnected components.  Indeed, for 
$k_0<0$ the
interaction is always attractive, but for $k_0>0$ it becomes repulsive at short distance. There
is thus a tendency for the particles to separate. Indeed, in the passive case $v_0=0$, there is phase transition
at $T=0$ for $k_0=0$, where the particles split into two packets located at $x=\pm \frac{1}{2}\sqrt{k_0/g}$ for $k_0>0$. At finite $T>0$, this transition becomes a change of behavior from a unimodal to a bimodal equilibrium density, which now occurs at a non-trivial value
\be
k_0 = 6 \,\frac{\Gamma(3/4)}{\Gamma(1/4)} \sqrt{g T} \;.
\ee
It is thus interesting to study the case of active noise to see whether similar phenomena occur.

\subsection{Main results}

Consider now the model of RTPs defined by \eqref{langevin_doublewell} with $T=0$ and $v_0>0$. 
In this case, there are two dimensionless parameters
\be \label{units} 
\tilde k_0 = \frac{k_0}{(g v_0^{2})^{1/3}} \quad , \quad \tilde \gamma = \frac{\gamma}{(g v_0^{2})^{1/3}} \;.
\ee
We perform a rescaling of the space and time variables as
\be 
x = (v_0/g)^{1/3} \tilde x \quad , \quad t = g^{-1/3} v_0^{-2/3} \tilde t \;,
\ee 
where $\tilde x$ and $\tilde t$ are now dimensionless. In the rest of the paper we will work in dimensionless coordinates but we will drop the tilde notation, which amounts to setting $g=1$ and $v_0=1$, so that the only independent parameters are $k_0$ and $\gamma$. 

As in the passive case, we will work in the reference frame of the center of mass (which, in the absence of external potential, diffuses freely as $\bar x \sim \sqrt{2 D_N t}$, with $D_N=\frac{1}{N}\frac{v_0^2}{2 \gamma}$). 

In the first part of the paper, Sec.~\ref{sec:sym}, we assume that the stationary density $\rho_s(x)$ is symmetric around $x=0$.
This can be realized by starting from an initial condition $\rho_s(x,t=0)$ which
is itself symmetric. Under this assumption, the odd moments of $\rho_s(x)$ vanish and 
the effective force field takes the form
\be \label{exprFtilde_intro}
\tilde F(x) = - \int dy (- k_0 (x-y) + (x-y)^3 ) \rho_s(y)  = - ( - k_0 + 3 m_2) x - x^3 = - x (x^2 - k) \;, 
\ee 
where we introduce
\be \label{def_k}
k = k_0 - 3 m_2 \quad , \quad m_n := \int dy \, y^n \rho_s(y)  \, ,
\ee 
and where we have defined the moments $m_n$ of the total density.
Here $k$ is the ``renormalized" version of the ``bare" interaction parameter $k_0$.
We see that the force acting on a given particle $\tilde F(x)$ only depends on the positions of the other particles through the second moment $m_2$ of the distribution $\rho_s(x)$. Hence we can compute the stationary distribution $\rho_s(x)$ for a given value of $k$ and compute a posteriori the corresponding value of the bare parameter $k_0$. Let us now summarize the main results.
We start by discussing the results in terms of the renormalized parameter $k$,
and discuss afterwards the relation between $k$ and $k_0$.

\begin{figure}[t]
\centering
\begin{minipage}[c]{0.4\linewidth}
\vspace{0.5cm}
\includegraphics[width=\linewidth,trim={0.2cm 0cm 0cm 0cm},clip]{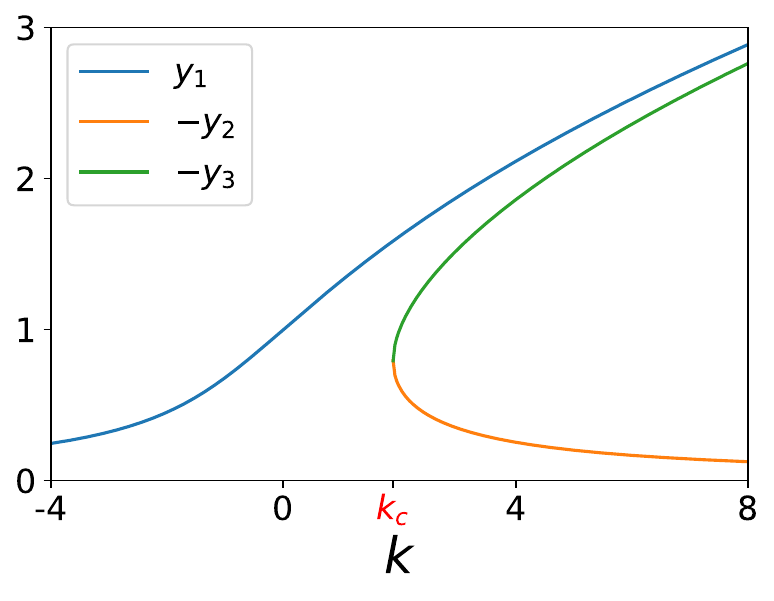}
\end{minipage}
\hspace{0.2cm}
\begin{minipage}[c]{0.55\linewidth}
\includegraphics[width=\linewidth,trim={5.6cm 0.2cm 7.6cm 0.2cm},clip]{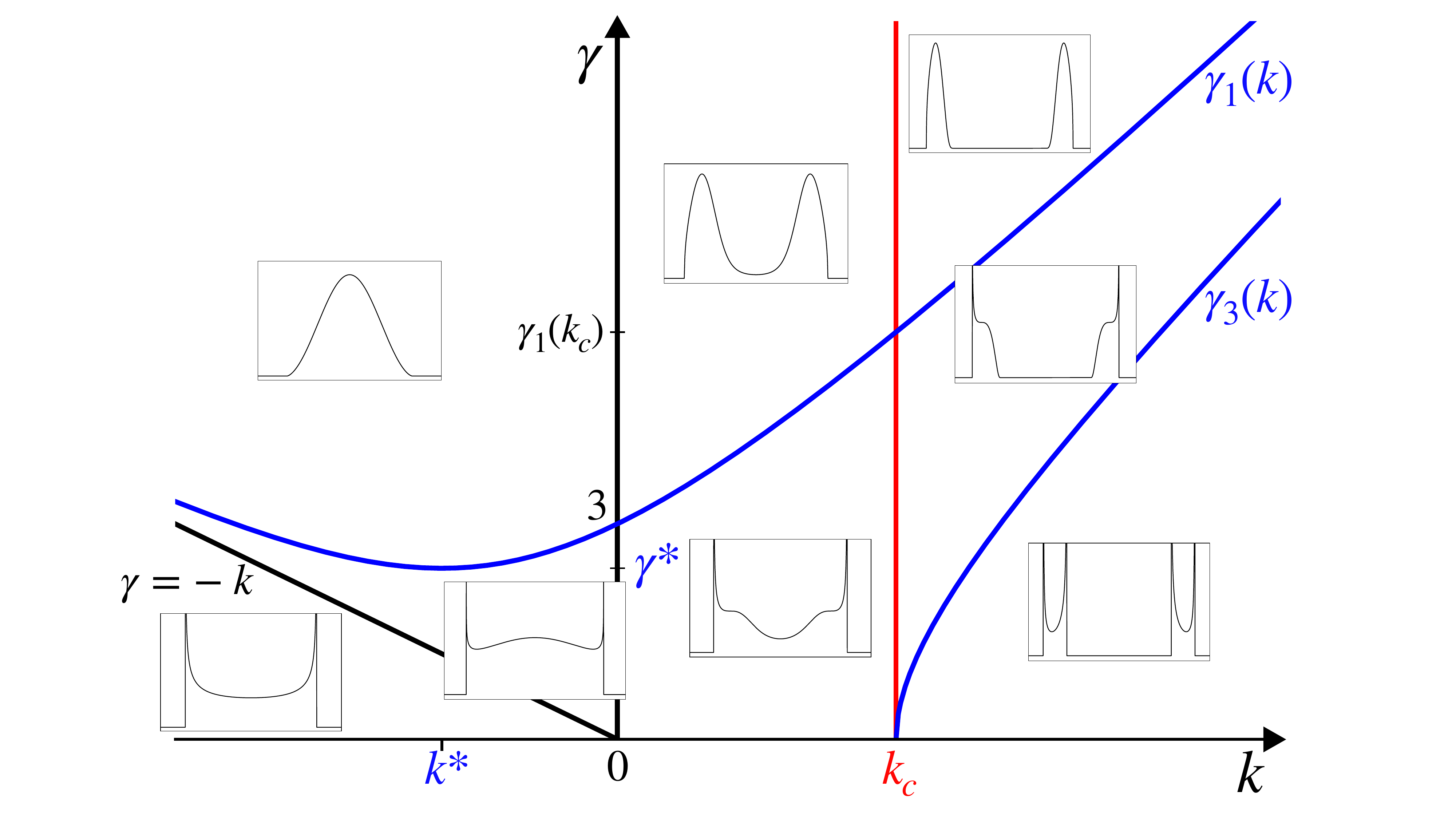}
\end{minipage}
\caption{Left: Plot of the absolute values of the real roots of  \eqref{cubic_eq}, $y_1$, $-y_2$ and $-y_3$, as a function of $k$. The largest root $y_1$ is real positive for any $k$ and is a continuous function of $k$. The roots $y_2$ and $y_3$ are real only for $k>k_c$ and are both negative, with $0<-y_2<-y_3<y_1$. Right: Diagram showing the different regimes in the behavior of the density $\rho_s(x)$, in the $(k,\gamma)$-plane. Each regime is illustrated by an inset plot of $\rho_s(x)$ obtained from the analytical expression for $N\to \infty$. The vertical red line at $k=k_c=3/2^{2/3}=1.88988...$ marks the transition from a connected support for $k<k_c$ to a disconnected support for $k>k_c$. The blue curves indicate a change in the edge behavior, from diverging to vanishing (at the exterior edges $\pm y_1$ for $\gamma_1(k)$ and at the interior edges $\pm y_3$ for $\gamma_3(k)$). The curve $\gamma_1(k)$ has a minimum at $k^*=-3/2^{4/3}=-1.19055...$, corresponding to $\gamma^*=3/2^{1/3}=2.38110...$, below which $\rho_s(x)$ always diverges at $\pm y_1$. Finally, the black lines $k=0$ and $k=-\gamma$ indicate a change of convexity of $\rho_s(x)$ around $x=0$. We also indicate the important intersection values $\gamma_1(0)=3$ and $\gamma_1(k_c)=9/2^{2/3}=5.66964...$~. The blue curve $\gamma_1(k)$ asymptotically coincides with the line $\gamma=-k$ as $k\to-\infty$, while for $k\to+\infty$, both $\gamma_1(k)$ and $\gamma_2(k)$ are asymptotically equal to $2k$. 
Note that this diagram is represented as a function of the renormalized parameter $k$, hence one should also take into account the mapping $k_0\to k$, which as discussed in the text is multi-valued in a small region (small $\gamma$ and $k$ near $k_c$).}
\label{phase_diagram_doublewell}
\end{figure}

The first result is that there is a transition in the support of the stationary density 
at a critical value $k=k_c$ of the renormalized parameter.
\begin{itemize}

\item For 
$k < k_c= 3/2^{2/3}$, the support of $\rho_s(x)$ is a single interval $[-y_1,y_1]$,
with $y_1>0$.
The explicit formula for $\rho_s(x)$ is given in Eq.~\eqref{rhos_joint}. 

\item For $k > k_c= 3/2^{2/3}$, the support of $\rho_s(x)$ is the union of two disjoint intervals $[-y_1,y_3]$
and $[-y_3,y_1]$, with $y_1>0$ and $y_3<0$. In that case the density has the form 
\be \label{rhos_disjoint_intro}
\rho_s(x) =  K (y_1^2-x^2)^{\eta_1-1} (x^2-y_2^2)^{\eta_2-1} (x^2-y_3^2)^{\eta_3-1} \, \theta(-y_3<|x|<y_1) \;,
\ee
where $K$ is a normalization constant and the exponents $\eta_i$
are given in \eqref{etavalues1}. In all cases $y_1$, $y_2$ and $y_3$ are roots 
of a cubic equation given in \eqref{cubic_eq}, and are plotted as a function of $k$ in Fig. \ref{phase_diagram_doublewell}
(left panel).

\item For $k=k_c$, i.e. in the critical case, the support of $\rho_s(x)$ is still the union of two disjoint intervals $[-y_1,y_3]$
and $[-y_3,y_1]$. What happens is that for $k>k_c$ the interval $[y_3,-y_3]$ is
filled by particles, but with a density which vanishes continuously as $k \to k_c^-$. 
For $k=k_c$ the stationary density reads
\be
\rho_s(x) = f_{\gamma/3k_c} \left( \sqrt{\frac{3}{k_c}} \, x \right) \quad , \quad f_\alpha(z) = A (4-z^2)^{\alpha-1} (z^2-1)^{-\alpha-2} e^{\frac{-6\alpha}{z^2-1}}
\, \theta(1<z^2<4) \;,
\ee
where $A$ is a normalization constant. From this one can deduce the critical value of $k_0$ which corresponds to $k=k_c=3/2^{2/3}$,
\be
k_{0,c} = k_c + 3 m_2(k_c) \quad , \quad m_2(k_c)= \frac{k_c}{3} \frac{ 
\int_{1}^{4} dz z^2 f_{\gamma/3k_c}(z) }{\int_{1}^{4} dz f_{\gamma/3k_c}(z)} \;.
\ee
\end{itemize}
In addition, we find that as $k\to k_c^-$, the density at $x=0$ vanishes continuously as
\be \label{rhosx0intro}
\rho_s(0) \propto \exp \left( -\frac{2^{1/3} \pi \gamma}{\sqrt{3(k_c-k)}} \right) \;,
\ee
i.e. it exhibits an essential singularity.

The transition between a connected and a disconnected support at $k=k_c$ is marked by a vertical red line in the diagram of Fig.~\ref{phase_diagram_doublewell} (right panel). This diagram also illustrates the different regimes in the plane $(k,\gamma)$ for the stationary density, depending on (i) its behavior near the edges $\pm y_1$ and $\pm y_3$, characterized by the two curves $\gamma_1(k)$ and $\gamma_3(k)$ (in blue in the figure), and (ii) its convexity around $x=0$. In particular, $\rho_s(x)$ vanishes algebraically at the exterior edges $\pm y_1$ when $\gamma>\gamma_1(k)$ and diverges when $\gamma<\gamma_1(k)$, and similarly at the interior edges when $k>k_c$ (with $\gamma_1(k)$ replaced by $\gamma_3(k)$).

In Fig.~\ref{phase_diagram_doublewell}, the behavior of the stationary density is represented as a function of $\gamma$ and of the renormalized interaction parameter $k$. The true physical parameter describing the interaction is however the bare parameter $k_0$. It is thus important to study the relation between $k$ and $k_0$, which can be computed a posteriori from \eqref{def_k}. This relation is plotted numerically in Fig.~\ref{m2vsk}. Crucially, we find that, while $k_0$ is a monotonously increasing function of $k$ for $\gamma>\gamma_c=0.1787...$, it becomes non-monotonous for $\gamma<\gamma_c$. This means that the function $k(k_0)$ is multi-valued on some range $[k_0^{min},k_0^{max}]$. In this range of values of $k_0$, the system admits two distinct stable stationary densities.

We conclude our study in Sec.~\ref{sec:asym} by investigating the possibility of non-symmetric steady states, for which $\rho_s(-x)\neq\rho_s(x)$. We find that such steady states are indeed possible in the regime where $k>k_c$, i.e. when the support admits two disconnected components. Indeed, in this case it is possible to observe steady states where one component contains more particles than the other, leading to an asymmetric density $\rho_s(x)$. Such steady states can be characterized by the value of the third moment $m_3$ (throughout the study we impose $m_1=0$, which amounts to fixing the center of mass at $x=0$). We find that $m_3$ may vary continuously inside a range of values $[-m_3^c,m_3^c]$ with $m_3^c=(k/k_c)^{3/2}-1$, implying an infinite number of possible steady states, depending on the initial condition.

{\blue Finally, a limit of interest for RTPs is the diffusive limit obtained by taking $\gamma, v_0 \to +\infty$ with $T_{\rm eff}=\frac{v_0^2}{2\gamma}$ fixed (at fixed $k_0$). 
In that limit one expects to recover the results for Brownian particles.
One can see from
\eqref{units} that it corresponds to $\tilde k_0 \to 0$, hence to
a support of the density which is always joint, as expected from Gibbs equilibrium at 
temperature $T_{\rm eff}$.}

In order to test these analytical predictions, obtained in the 
limit $N \to +\infty$, we have performed direct numerical simulations of
the equation of motion \eqref{langevin_doublewell}. The stationary density is determined
by averaging over a large time window. We find a good agreement between theory and numerics already for $N=100$ (see Fig.~\ref{fig_density_symmetric}). In particular, both the bistability (see Fig.~\ref{fig_bistable}) and the existence of asymmetric steady states (see Fig.~\ref{fig_asym}) are confirmed by these numerical results. The case of small values of $N$ is discussed in Appendix~\ref{app:N2}.



%

\section{Derivation of the results in the symmetric case} \label{sec:sym}


We now derive the results announced in the previous section for the model defined in \eqref{langevin_doublewell}, with the double-well interaction potential defined in \eqref{doublewell}.

\subsection{Passive case} \label{sec:brownian}

Let us first discuss the purely passive case, i.e. $v_0=0$ and $T>0$, for which the system reaches canonical equilibrium. The distribution of the 
particle positions is then given by the Gibbs measure
\be
{\cal P}(x_1,...,x_N)  \propto e^{- \frac{1}{N T} \sum_{i<j} W(x_i-x_j) } \;.
\ee
for any (sufficiently attractive) interaction potential $W$ such that ${\cal P}$ is normalizable on
the real axis. 
In the limit of large $N$ one can describe the system by a density field $\rho(x,t)=\frac{1}{N}\sum_i \delta(x-x_i)$,
with a probability distribution (see e.g. \cite{DeanMajumdar2008})
\be 
{\cal P}[\rho] 
\propto \int d\lambda e^{- \frac{N}{T} ( \frac{1}{2} \int dx dx' \rho(x) \rho(x') W(x-x') + T  \int dx \rho(x) \log \rho(x)  + i \lambda (\int \rho(x) -1))  } \;,
\ee 
where the entropy term (second term) arises from the Jacobian of the mapping between the two descriptions, 
and $i \lambda$ is a Lagrange multiplier enforcing the constraint $\int dx \rho(x)=1$. 
In the large $N$ limit the functional integral is dominated by a saddle point. Taking a functional derivative with respect
to the density of the term in the exponential, we obtain that the optimal density satisfies
\be \label{eq_doublewell_passive}
\rho_{eq}(x) = K e^{- \frac{1}{T} \int dx' \rho_{eq}(x') W(x-x')} \;,
\ee 
where $K$ is a normalization constant. This is a self-consistent equation for the equilibrium density $\rho_{eq}(x)$. One can check that it coincides with the equation \eqref{self_intro} for the purely active case in the limit $v_0,\gamma \to +\infty$ with $T=T_a = v_0/(2 \gamma)$.

Let us now specialize to the double-well interaction potential $W(x) = - k_0 \frac{x^2}{2} + g \frac{x^4}{4}$ with $g>0$ so that the equilibrium state is a bound state,
and define the moments
\be 
m_n= \int dx \, x^n \rho_{eq}(x) \;.
\ee 
Since the problem is invariant by translation we can work in the reference frame of the center of mass, such that $m_1=0$.
Then one has
\be
\int dx' \rho_{eq}(x') W(x-x') = - \frac{k_0}{2} (x^2 + m_2) + \frac{g}{4} ( x^4 + 6 m_2 x^2 - 3 m_3 x + m_4) \;.
\ee 
Since the thermal noise is unbounded the support of the equilibrium density for $T>0$ is a single interval. Given the symmetries of the problem, the equilibrium density should then be even in $x$ (below we will see however that in the purely active case, the support can be disjoint, allowing for asymmetric stationary states). We can thus fix $m_3=0$, leading to
\be 
\rho_{eq}(x) = K'  e^{- \frac{1}{T} ( -\frac{k}{2} x^2 + g \frac{x^4}{4} ) } \quad , \quad k = k_0 - 3g m_2  \;,
\ee 
where $K'$ is another normalization constant. 
In that case the equilibrium density is thus simply related to the interaction potential $W(x)$, up to a renormalization
of the parameter $k_0 \to k$, which should be determined self-consistently. 

For $T \to 0$ there is a phase transition at $k=0$, from a single delta peak at $x=0$ for $k<0$, to two delta peaks at $x=\pm \sqrt{k/g}$ for $k>0$.
The self-consistency condition gives, for $k_0 >0$ and $T=0$,
\be 
m_2=\frac{k}{g} \quad , \quad k = \frac{k_0}{4} \;,
\ee 
and $k=k_0$ for $k_0<0$. Hence the transition occurs at $k_0=0$. This can also be obtained directly at $T=0^+$
from the minimization of the interaction energy $E_{\rm int} = \frac{1}{2} \int dx dx' \rho(x) \rho(x') W(x-x')$
(since the entropy term vanishes in this limit),
with the ansatz $\rho_{\rm eq}(x)=p \delta(x-x_1) + (1-p) \delta(x-x_2)$. Indeed, in that
case one has $E_{\rm int} = p(1-p) [-k_0(x_1-x_2)^2+\frac{g}{2}(x_1-x_2)^4] $.
For $k_0<0$, the minimum is $x_1=x_2=0$. For $k_0>0$, the minimum is such that $E_{\rm int}<0$, thus to minimize the energy one has to take $p=1/2$. By symmetry, we must then have $x_1=-x_2$, from which we deduce that the optimal value is $x_{1,2}=\pm\frac{1}{2}\sqrt{k_0/g}$, which coincides with the solution we have just found.

For $T>0$, the phase transition disappears, but at 
any $T$ there is a change of behavior from a unimodal to a bimodal density at $k=0$, i.e. at 
\be
k_0 = 3g m_2(0) \quad , \quad m_2(0) = \frac{\int dx \, x^2 e^{-\frac{g}{T}\frac{x^4}{4}}}{\int dx \, e^{-\frac{g}{T}\frac{x^4}{4}}} = 2 \frac{\Gamma(3/4)}{\Gamma(1/4)} \sqrt{\frac{T}{g}} \;.
\ee
This change of behavior is characterized by the change of sign of the second order derivative of the density $\rho_{eq}(x)$ at $x=0$.

\subsection{Calculation of the stationary density in the purely active case} \label{sec:rho_derivation}

\begin{figure}
\centering
\includegraphics[width=0.32\linewidth,trim={2cm 0.5cm 1.95cm 0.5cm},clip]{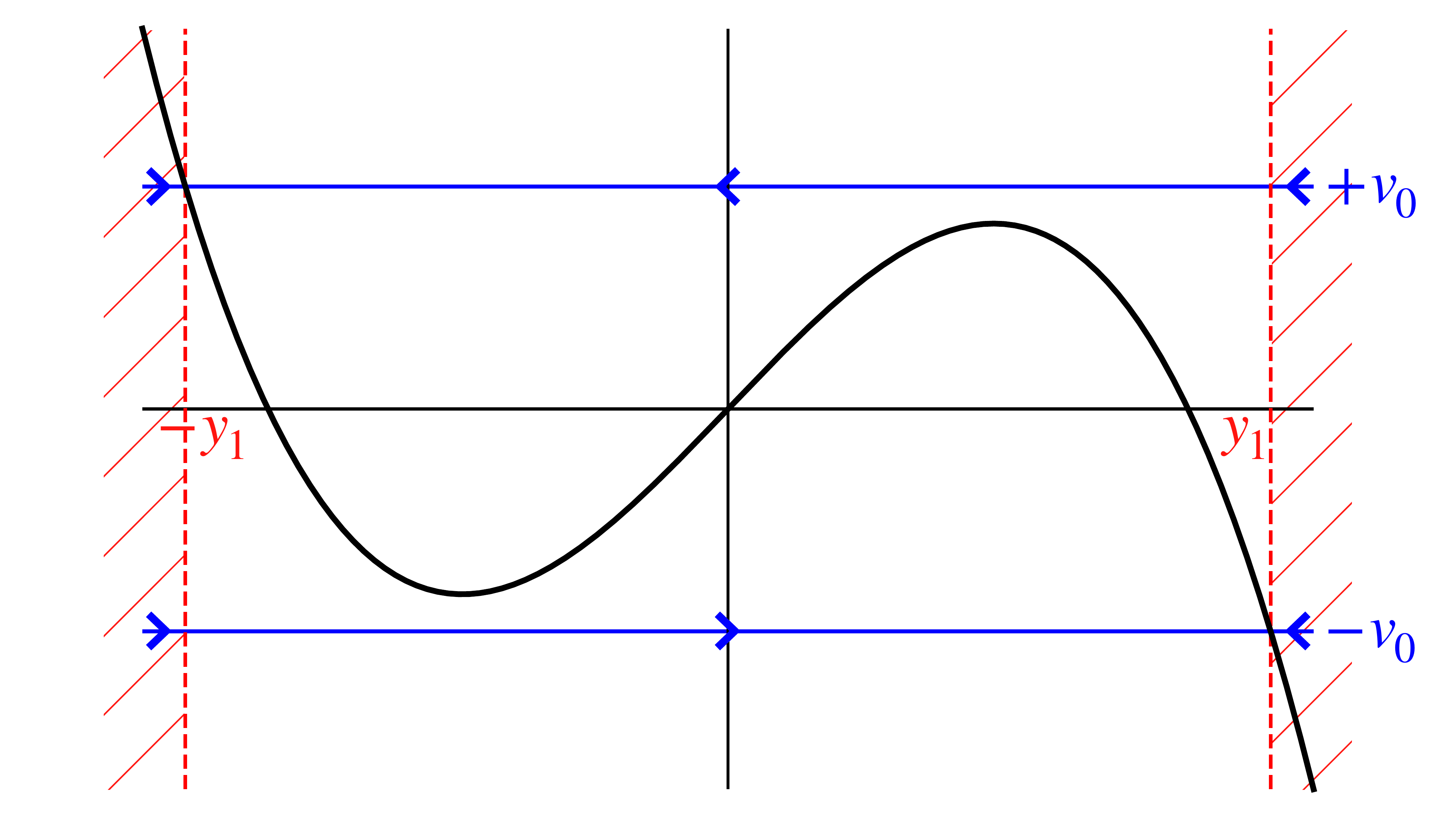}
\includegraphics[width=0.32\linewidth,trim={2cm 0.5cm 1.95cm 0.5cm},clip]{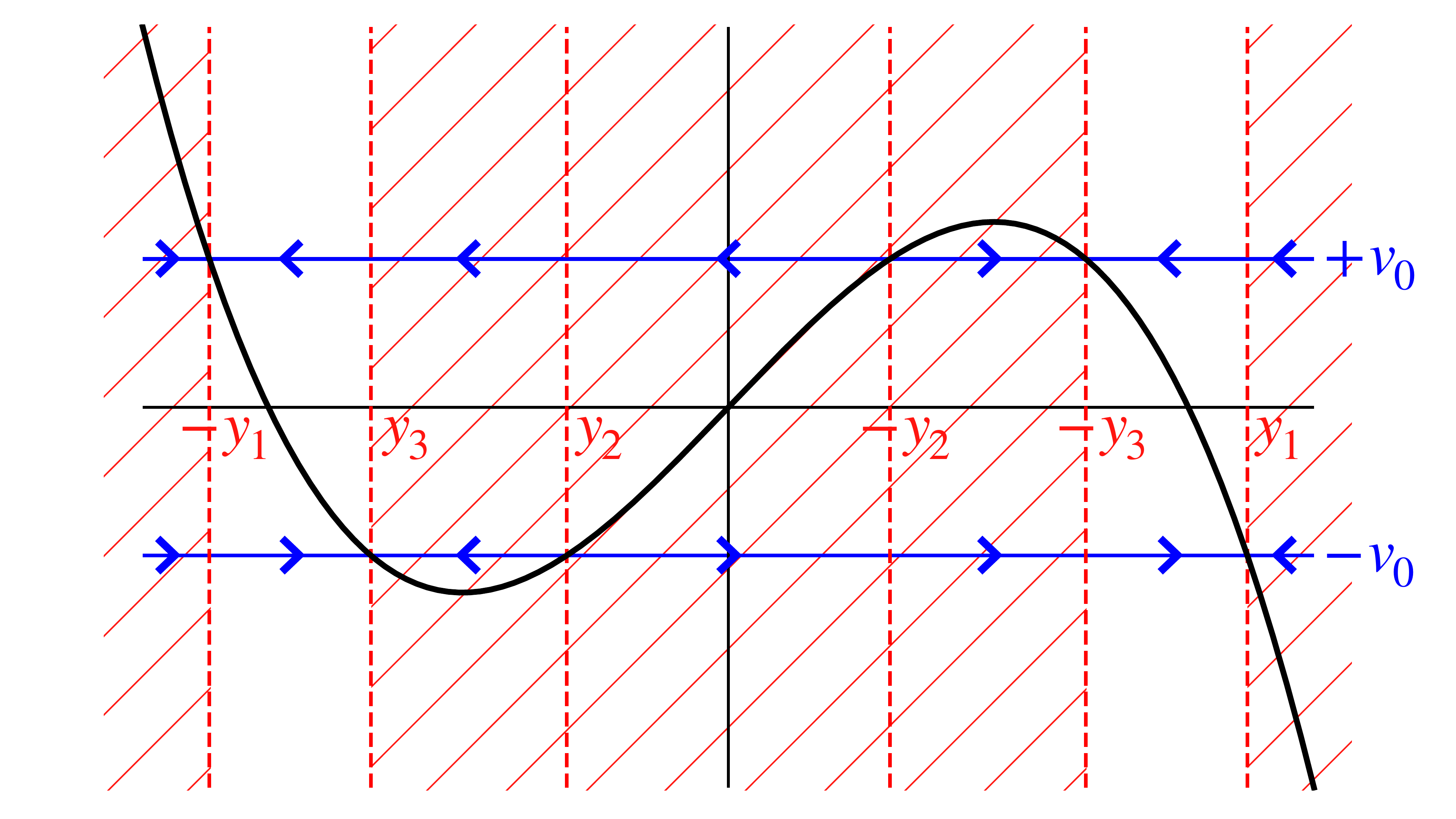}
\includegraphics[width=0.32\linewidth,trim={2cm 0.5cm 1.95cm 0.5cm},clip]{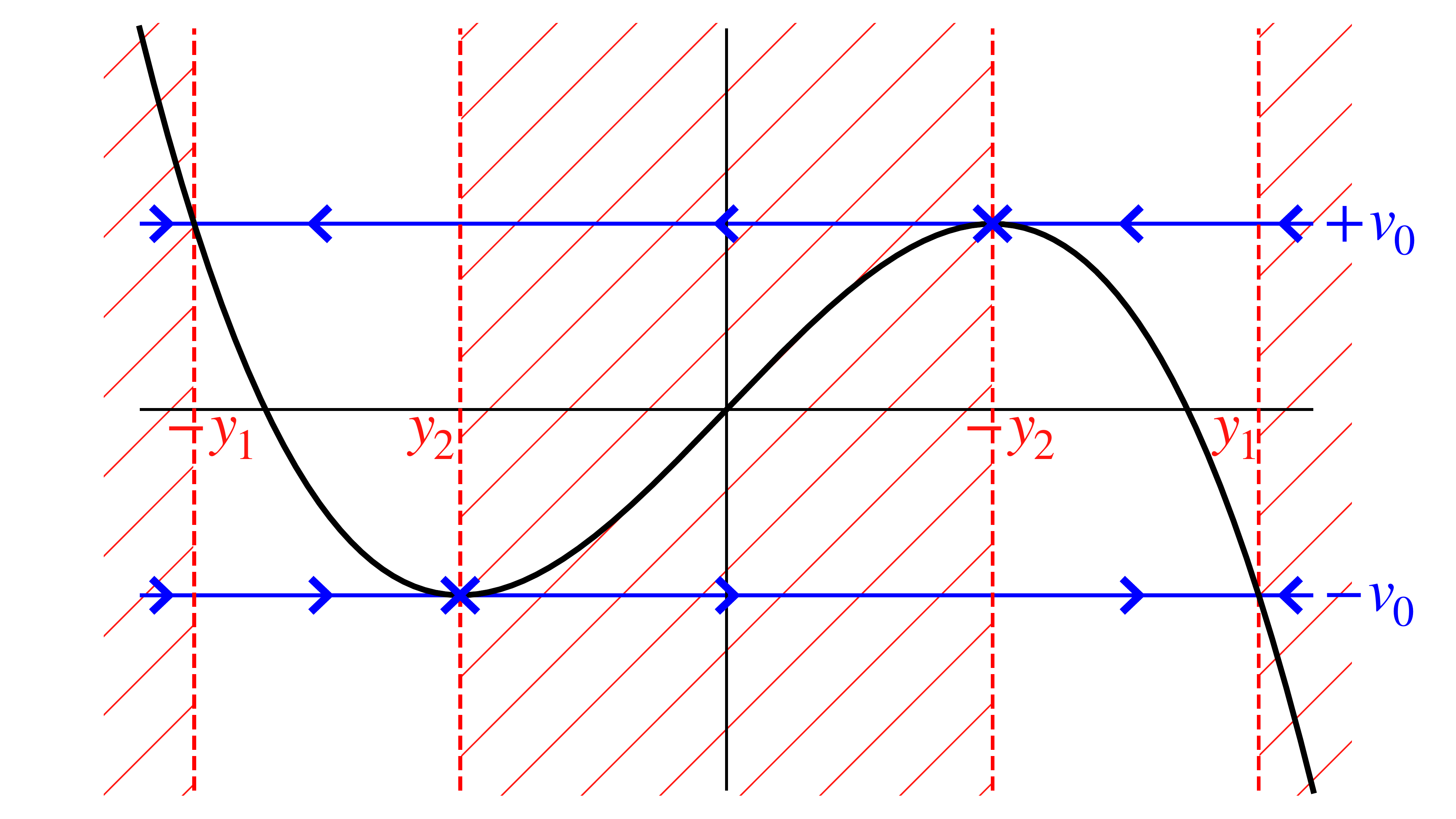}
\caption{Plot of $\tilde F(x)$ (black curve) in the 3 different regimes: $\Delta<0$ (left), $\Delta>0$ (center) and $\Delta=0$ (right). The regions hatched in red are inaccessible to the particles in the stationary state. Particles in the $+$ state move to the right when the curve is above the line $-v_0$ (in blue), and to the left when it is below. Particles in the $-$ state move to the right when the curve is above the line $+v_0$ (in blue), and to the left when it is below. 
We recall that the particles switch between the states $+$ and $-$ with rate $\gamma$. 
The arrows on the bottom blue line indicate the direction of motion of $+$ particles, while the arrows on the top blue line indicate the direction of motion of $-$ particles. Crosses indicate points where the total velocity vanishes without changing sign, meaning that the particle would take an infinite time to reach this point (thus a tumbling event always occurs before it reaches it).}
\label{stability_diagram}
\end{figure}

We now focus on the purely active case $T=0$ and $v_0>0$. For now we assume that the stationary density $\rho_s(x)$ is even in $x$. We start from the self consistent equation \eqref{self_intro}, which we rewrite as
\be \label{rhos_formal}
\rho_s(x) = \frac{K}{(1-\tilde F(x))(1+\tilde F(x))}
\exp \left( \gamma \int^x dy \frac{1}{1-\tilde F(y)} - \gamma \int^x dy \frac{1}{1+\tilde F(y)} \right) \;, 
\ee 
where $K$ is a normalization constant (the lower limits of the integrals are irrelevant due to normalization but should be chosen to avoid divergences, as we discuss below), and we recall that $\tilde F(x)$ was expressed in \eqref{exprFtilde_intro} as a function of the renormalized interaction parameter $k$. To determine the support of $\rho_s(x)$ we need to study the roots of the cubic equations
(i.e. of the denominators in \eqref{rhos_formal}) 
\bea \label{cubic_eq}
&& 0 = -1 - \tilde F(y) = y^3 - k y - 1 = (y-y_1)(y-y_2)(y-y_3) \;, \\
&& 0 = 1 - \tilde F(y) = y^3 - k y + 1  = (y+y_1)(y+y_2)(y+y_3) \;, \nn
\eea 
where we note that $y_1+y_2+y_3=0$. 
The discriminant of a depressed cubic equation $ t^{3}+pt+q=0$ is defined as
$\Delta=-(4p^3+27q^2)$. For both equations in \eqref{cubic_eq} this gives
\be 
\Delta = 4k^3 - 27 \;.
\ee 
This is negative for small (or negative) values of $k$, but it becomes positive when $k$ crosses the critical value $k_c=3/2^{2/3}$. There are thus 3 distinct cases, depending on the value of $k$, as represented schematically in Fig. \ref{stability_diagram} where we also describe the dynamics leading to these cases:
\begin{itemize}
    \item If $k<k_c$, i.e. $\Delta<0$, there is only one real root $y_1$. In this case the support of the stationary density $\rho_s(x)$ is simply given by $[-y_1,y_1]$.
    \item If $k>k_c$, i.e. $\Delta>0$, there are 3 real roots $y_1>0>y_2>y_3$. The support of the density is the union of two disjoint intervals $[-y_1,y_3]$ and $[-y_3,y_1]$.
    \item In the marginal case $k=k_c$, i.e. $\Delta=0$, there is one positive real root $y_1$ and one real negative double root $y_2=y_3$. In this case the support is again the union of $[-y_1,y_3]$ and $[-y_3,y_1]$, but the density vanishes exponentially at $\pm y_3$ instead of obeying a power law as in the other cases (see below).
\end{itemize}
We now compute the stationary density explicitly in these 3 cases. In the next section we will discuss in more detail its main features.
\\

{\bf Disjoint support.} We start with the case $\Delta>0$, i.e. $k>k_c=3/2^{2/3}$, where there are 3 real roots given by 
\be \label{yvalues1}
y_{n+1} = 2\sqrt{\frac{k}{3}} \cos \left( \frac{1}{3} \arccos \left(\left(\frac{k_c}{k}\right)^{3/2}\right) - \frac{2\pi n}{3}\right) \quad , \quad n=0,1,2 \;.
\ee
One has $y_1>-y_3>-y_2>0$.
Let us start by determining the support of the stationary density, using the diagram of Fig.~\ref{stability_diagram} (central panel). First, all the particles which are outside the interval $[-y_1,y_1]$ have a total velocity $\tilde F(x) + \sigma$ which drives them towards this interval, so that at large times all the particles are inside the interval $[-y_1,y_1]$.
In addition, all the particles that are inside the interval $[-y_2,-y_3]$ (resp. $[y_3,y_2]$) move towards the right (resp. left) and become trapped inside the interval $[-y_3,y_1]$ (resp. $[-y_1,y_3]$) after some time. Since nothing prevents the particles initially located inside the central interval $[y_2,-y_2]$ to escape, it will empty itself as time goes on and in the stationary state all the particles will be located in one of the two intervals $[-y_1,y_3]$ and $[-y_3,y_1]$. In addition, if we assume that the initial distribution of the particles is symmetric, there will be on average the same number of particles in these two intervals, and the stationary distribution will be symmetric, compatible with our assumption. However, if there is an asymmetry in the initial distribution, the stationary state may be asymmetric. We will discuss this case in Sec.~\ref{sec:asym}

Thus in the present case $k>k_c$, the stationary density has two disjoint supports $[-y_1,y_3]$ and $[-y_3,y_1]$. Using \eqref{rhos_formal}, along with
\be
\int^x dy \frac{1}{1-\tilde F(y)} -  \int^x dy \frac{1}{1+\tilde F(y)} 
= \frac{ \log(y_1^2-x^2) }{(y_1-y_2)(y_1-y_3)} + \frac{ \log(x^2-y_2^2) }{(y_2-y_1)(y_2-y_3)} + \frac{ \log(x^2-y_3^2) }{(y_3-y_1)(y_3-y_2)} 
\ee
(we recall that inside the support one has $y_2^2<y_3^2<x^2<y_1^2$), we obtain
\be \label{rhos_disjoint}
\rho_s(x) =  K (y_1^2-x^2)^{\eta_1-1} (x^2-y_2^2)^{\eta_2-1} (x^2-y_3^2)^{\eta_3-1} \;,
\ee
where the three exponents are 
\bea \label{etavalues1}
&& \eta_1 = \frac{\gamma}{(y_1-y_2)(y_1-y_3)} = \frac{\gamma}{3 y_1^2-k}  >0 \quad , \quad \eta_2 = \frac{\gamma}{(y_2-y_1)(y_2-y_3)} =  \frac{\gamma}{3 y_2^2-k} <0 \;, \nonumber \\
&&  \eta_3 = \frac{\gamma}{(y_3-y_1)(y_3-y_2)}  = \frac{\gamma}{3 y_3^2-k} >0 \quad , \quad \eta_1+\eta_2+\eta_3=0 \;.
\eea
Here we have used that $-\tilde F'(y)=(y-y_1)(y-y_2)+(y-y_1)(y-y_3)+(y-y_2)(y-y_3) = 3y^2-k$ to rewrite the exponents $\eta_i$.
Note that since the density is zero around $y_2$, the exponent $\eta_2-1$ does not correspond to any edge behavior. Thus we observe a change of behaviour between a divergence ($\eta_{1,3}<1$) and a vanishing density ($\eta_{1,3}>1$) at all edges, but it will occur for different values of the parameters for the edges at $\pm y_1$ and at $\pm y_3$. Note that since $y_3^2<y_1^2$, one has $\eta_3>\eta_1$, and thus as $\gamma$ increases the vanishing of the density occurs first at the edges $\pm y_3$ (we recall that the $y_i$ only depend on $k$). Finally, the value of the normalization constant $K$ is obtained by normalizing $\rho_s(x)$ to 1, and the value of $k_0$ corresponding to this density profile can be computed from \eqref{def_k}. One obtains the relation
between $k_0$ and $k$
\\
\be \label{k0versusk1}
k_0 = k + 3 m_2(k) \quad , \quad m_2(k) = \frac{ \int_{-y_3}^{y_1} dx x^2 (y_1^2-x^2)^{\eta_1-1} (x^2-y_2^2)^{\eta_2-1} (x^2-y_3^2)^{\eta_3-1} }{
\int_{-y_3}^{y_1} dx (y_1^2-x^2)^{\eta_1-1} (x^2-y_2^2)^{\eta_2-1} (x^2-y_3^2)^{\eta_3-1}} \;.
\ee 
where the $y_i$ and $\eta_i$ are given
in \eqref{yvalues1} and \eqref{etavalues1}
as a function of $k$. 
This relation will be studied in more detail in Section \ref{sec:bistability}.
For most of the parameter range it is monotonous and single-valued, with however
an interesting small region where a bistability occurs. 
\\

{\bf Critical case.} We now consider the critical case $\Delta=0$, i.e. $k=k_c=3/2^{2/3}$. Then there are one simple and one double real root,
\be 
y_1 = \frac{3}{k_c} = 2^{2/3} \quad , \quad y_2=y_3 = - \frac{3}{2 k_c} = - 2^{-1/3} \;.
\ee
This gives
\bea
-1-\tilde F(y) &=& (y - 2^{2/3})  (y + 2^{-1/3})^2 \;, \\
1 - \tilde F(y) &=& (y + 2^{2/3})  (y - 2^{-1/3})^2 \;.
\eea
Once again, we start by determining the support of the stationary density, using the diagram of Fig.~\ref{stability_diagram} (right panel). As in the previous case, the particles which are outside the interval $[-y_1,y_1]$ all move towards this interval, so that at large times all the particles are inside this interval. In addition, the particles can escape the interval $[y_2,-y_2]$ in a finite time, but they cannot go back inside this interval since it takes an infinite time to cross the lines $\pm y_2$ from the outside (the total velocity is either directed away from the interval or vanishes at this point). Thus, in the stationary state all the particles are located inside $[-y_1,y_2]\cup[-y_2,y_1]$.




Inside the disjoint support, the density reads, using again \eqref{rhos_formal},
\bea 
\rho_s(x) &=& \frac{K}{(2^{4/3}-x^2)  (x^2 - 2^{-2/3})^2} \exp \left( -\frac{2}{3} \frac{\gamma}{x^2-2^{-2/3}} + \frac{2^{2/3}\gamma}{9} \ln \big(\frac{2^{4/3}-x^2}{x^2-2^{-2/3}} \big) \right) \\
&=& K (2^{4/3}-x^2)^{\frac{2^{2/3}}{9}\gamma-1}  (x^2 - 2^{-2/3})^{-\frac{2^{2/3}}{9}\gamma-2} \exp \left( -\frac{2}{3} \frac{\gamma}{x^2-2^{-2/3}} \right) \;,
\eea 
where we recall that the support is $2^{-2/3} < x^2 < 2^{4/3}$.
Contrary to the previous case, the density always vanishes exponentially at $\mp y_2=\pm 2^{-1/3}$. At $\pm y_1=\pm2^{2/3}$, it vanishes algebraically for $\gamma>\gamma_s=9/2^{2/3}$ and it diverges for $\gamma<\gamma_s$. Once again, the normalization constant $K$ and the parameter $k_0$ can be computed a posteriori from the expression of $\rho_s(x)$. In particular the critical value of $k_0$ is given by (recalling $k_c=3/2^{2/3}$)
\be \label{k0versusk_critical}
k_{0,c} = \frac{3}{2^{2/3}} + 3 m_2(k_c) \;, \quad m_2(k_c) = \frac{ \int_{2^{-1/3}}^{2^{2/3}} dx x^2 (2^{4/3}-x^2)^{\frac{2^{2/3}}{9}\gamma-1}  (x^2 - 2^{-2/3})^{-\frac{2^{2/3}}{9}\gamma-2} \exp \left( -\frac{2}{3} \frac{\gamma}{x^2-2^{-2/3}} \right) }{
\int_{2^{-1/3}}^{2^{2/3}} dx (2^{4/3}-x^2)^{\frac{2^{2/3}}{9}\gamma-1}  (x^2 - 2^{-2/3})^{-\frac{2^{2/3}}{9}\gamma-2} \exp \left( -\frac{2}{3} \frac{\gamma}{x^2-2^{-2/3}} \right)} \;.
\ee 
\\

{\bf Connected support.} We now consider the last case, $\Delta<0$, i.e. $k<k_c$, for which there is only one real root
\be \label{defy1_joint}
y_1 = u_+ + u_- \quad , \quad u_\pm = \frac{1}{2^{1/3}} \left( 1 \pm \sqrt{1-\left( \frac{k}{k_c} \right)^3} \right)^{1/3} \quad , \quad u_+ u_- = \frac{k}{3} \;,
\ee
with $u_+ u_- = \frac{1}{2^{2/3}} \frac{k}{k_c}=k/3$, and two complex conjugate roots
\be \label{defy23_joint}
y_2 = \bar y_3 = \omega u_+ + \bar \omega u_- \quad , \quad \omega = \frac{-1+i\sqrt{3}}{2} \;,
\ee
such that 
\bea \label{ab} 
(y \pm y_2)(y \pm y_3) = y^2 \pm a y + b \;, \quad \text{with} \quad
a &=& 2 {\rm Re}(y_2) = -(u_+ + u_-) = -y_1 \;,  \\
\text{and} \quad b &=& |y_2|^2 = u_+^2 + u_-^2 - u_+u_- = y_1^2 - k \;. \nn
\eea
In this case the support of the stationary density is a single interval, $[-y_1,y_1]$ -- see Fig.~\ref{stability_diagram} (left panel). We now have to compute
\bea 
\int^x dy \frac{1}{1-\tilde F(y)} -  \int^x dy \frac{1}{1+\tilde F(y)} 
&=& \int^x dy \frac{1}{(y+y_1)(y^2-y_1 y+b)} +  \int^x dy \frac{1}{(y-y_1)(y^2+y_1 y+b)} \nn \\
&=& \frac{1}{b+2y_1^2} \left[ \ln (y_1^2-x^2)  - \frac{1}{2}\ln \left((x^2+b)^2 - y_1^2 x^2 \right) \nn \right.\\
&&\quad \left. - \frac{3 y_1}{\sqrt{4b-y_1^2}} \left( \arctan \left(\frac{y_1-2x}{\sqrt{4b-y_1^2}}\right) + \arctan \left( \frac{y_1+2x}{\sqrt{4b-y_1^2}} \right) \right) \right] \nn \;. \\
\eea
Here we have used that
\be 
\frac{1}{(y+y_1)(y^2-y_1 y+b)} = \frac{1}{b + 2 y_1^2} (\frac{1}{y+y_1} - \frac{y - 2y_1}{y^2-y_1 y+b} ) \;.
\ee 
Using again \eqref{rhos_formal}, we obtain the stationary density as
\bea \label{rhos_joint}
\rho_s(x) 
 &=& K (y_1^2-x^2)^{\frac{\gamma}{3y_1^2-k}-1} \left((x^2+y_1^2-k)^2 - y_1^2 x^2 \right)^{-\frac{\gamma}{2(3y_1^2-k)}-1} \nn \\
&& \quad \quad \times \exp \left( -\frac{3 \gamma y_1}{(3y_1^2-k)\sqrt{3y_1^2-4k}} \left( \arctan \left(\frac{y_1-2x}{\sqrt{3y_1^2-4k}}\right) + \arctan \left( \frac{y_1+2x}{\sqrt{3y_1^2-4k}} \right) \right) \right)
\;,
\eea
where we recall that $y_1$ 
is given in 
\eqref{defy1_joint}. 
One can check that the factor $(x^2+y_1^2-k)^2 - y_1^2 x^2$ is always strictly positive for $x$ in the support. Indeed,
for $x>0$ the condition $x^2- y_1 x + y_1^2-k>0$ is always satisfied (the discriminant is $\Delta=4k-3y_1^2=-3(y_1^2-4u_+u_-)=-3(u_+-u_-)^2<0$). As in the previous cases, the constant $K$ 
is obtained from the normalization condition $\int_{-y_1}^{y_1} dx \rho_s(x) = 1$. One can then compute the second moment $m_2(k)=\int_{-y_1}^{y_1} dx \, x^2 \rho_s(x)$ from 
\eqref{rhos_joint},
and deduce the corresponding value of $k_0=k+3 m_2(k)$.


Once again, we find that as $\gamma$ is increased there is a change of behavior between a diverging and a vanishing density at the two edges $\pm y_1$, which occurs at the value $\gamma=3y_1^2-k$, as in the case of disjoint support for the external edges. 

\begin{figure}
\centering
\includegraphics[width=0.45\linewidth]{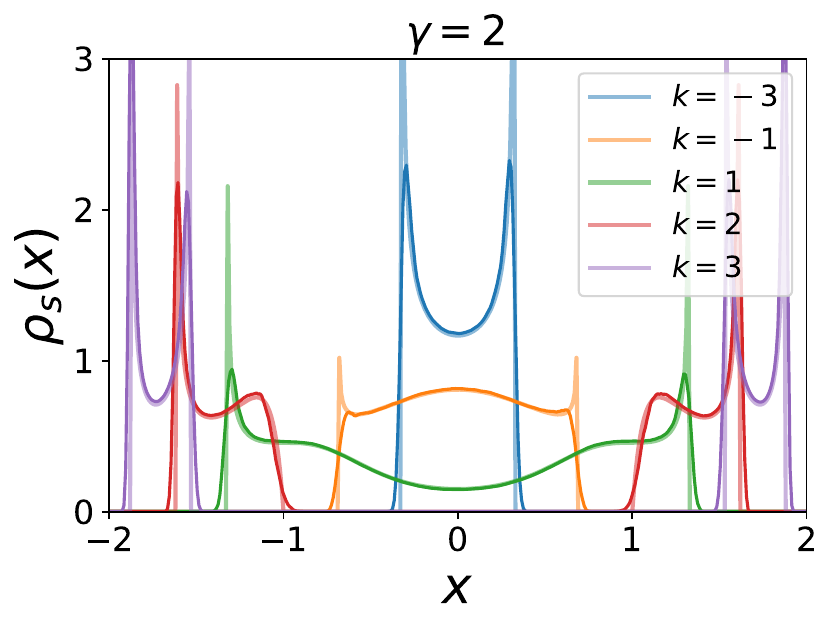}
\includegraphics[width=0.45\linewidth]{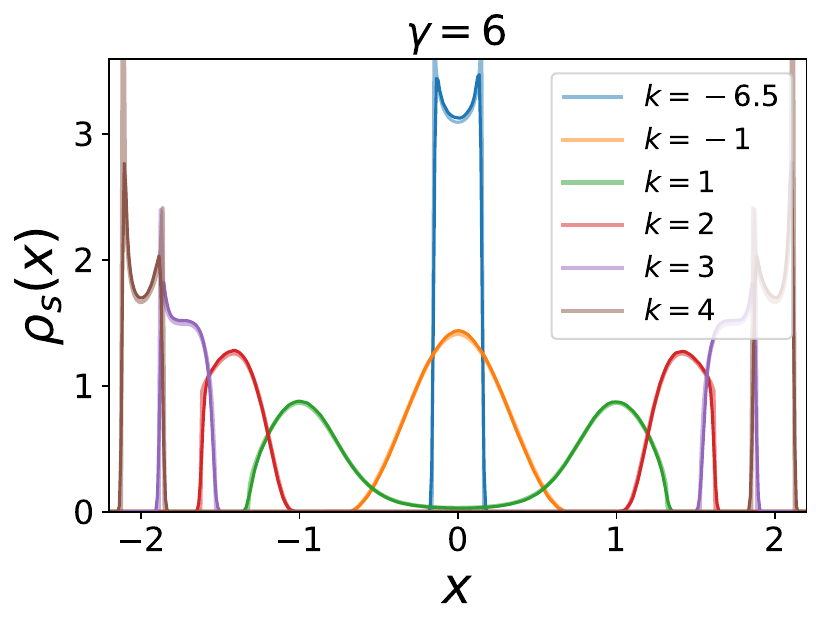}
\caption{Plots of the density $\rho_s(x)$ for different values of $k$ (corresponding to different regimes in the diagram of Fig.~\ref{phase_diagram_doublewell} and for $\gamma=2$ (left) and $\gamma=6$ (right) (with $g=1$, $v_0=1$ and $x_{init}=1$). The plots in light colors show the theoretical prediction (given by \eqref{rhos_disjoint} or \eqref{rhos_joint}) while the darker lines were obtained by simulating the stochastic dynamics for $N=100$ particles and averaging the histogram of positions in the steady state. For each value of $k$, the corresponding value of $k_0$ to be used in the simulations was computed numerically using the relation \eqref{def_k}, {\blue and is given in Appendix~\ref{app:table}}.}
\label{fig_density_symmetric}
\end{figure}

Finally, we have tested these predictions using numerical simulations. We solve numerically the equation
of motion
\eqref{langevin_doublewell} at $T=0$ for $N=100$ particles, starting from a uniform initial condition on some interval $[-x_{init},x_{init}]$
and we determine the stationary total density by time averaging the density in the center of mass frame at large time. For $k>k_c$, we instead start from an initial condition where the particles are separated into two groups of equal size, which allows to obtain a symmetric stationary state. Different choices of initial condition may lead to an asymmetric stationary state, see Sec.~\ref{sec:asym}. The results are displayed in Fig.~\ref{fig_density_symmetric}
and show excellent agreement {\blue (although finite $N$ effects are visible near the edges of the support in some cases).}

\subsection{General discussion of the results}

In this section, we analyze the expressions of the stationary total density $\rho_s(x)$ obtained in the 
previous subsection, and study how the general features of the density evolve depending on the parameters. 
We recall that here we restrict to the case where the stationary density is even in $x$.

\subsubsection{Relation between $k$ and $k_0$}

\begin{figure}
    \centering
    \hspace{0.15cm}
    \includegraphics[width=0.45\linewidth]{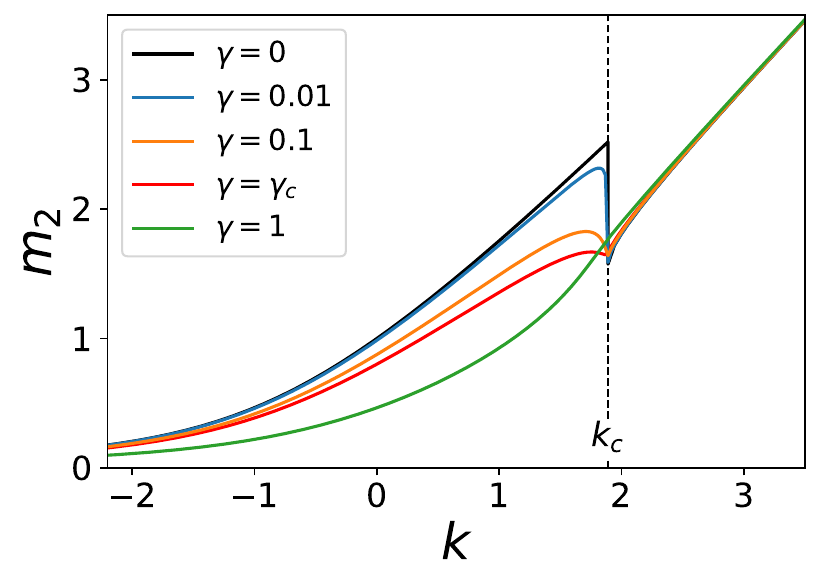}
    \hspace{0.1cm}
    \includegraphics[width=0.487\linewidth]{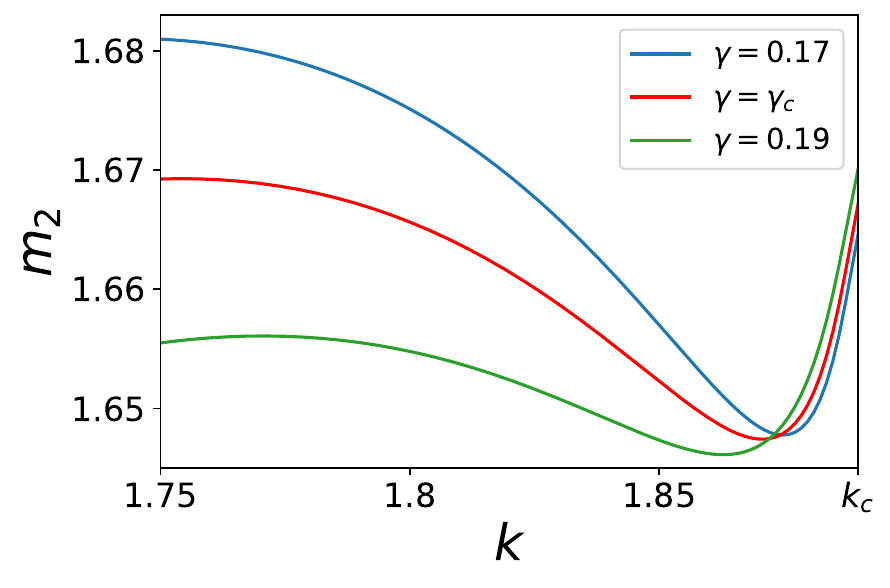}
    \includegraphics[width=0.475\linewidth]{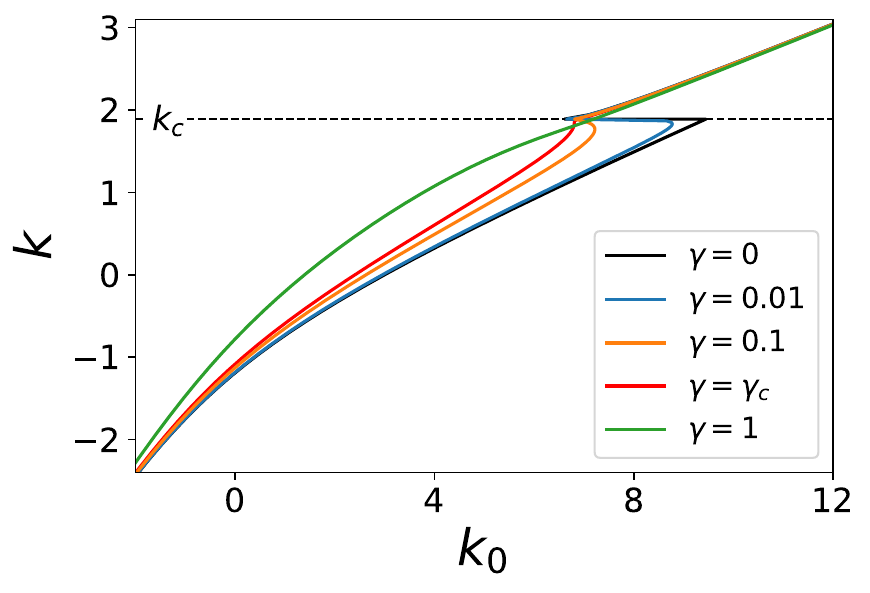}
    \includegraphics[width=0.48\linewidth]{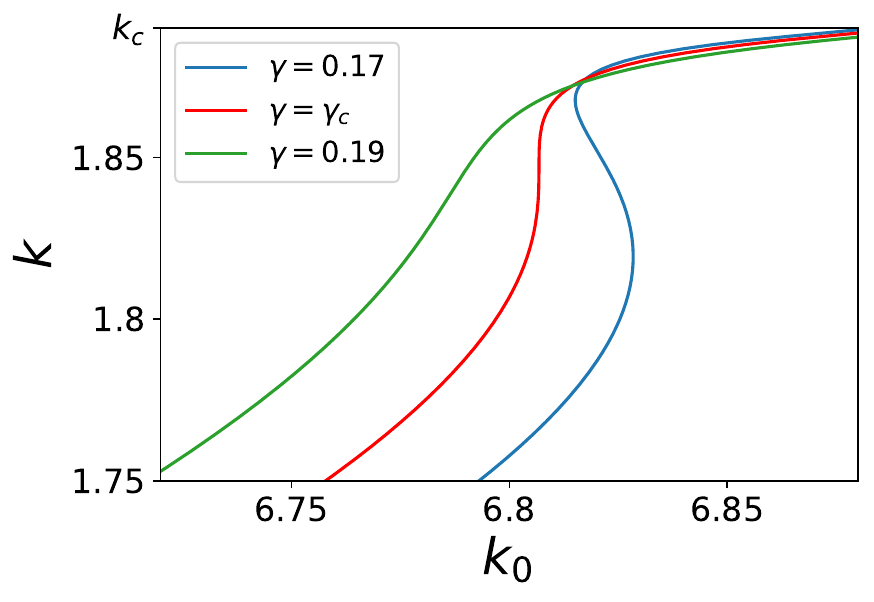}
    \caption{Top left: Plot of $m_2$ versus $k$ for different values of $\gamma$, {\blue obtained by evaluating numerically the integral in \eqref{k0versusk1} for $k>k_c$, and from an equivalent expression deduced from \eqref{rhos_joint} for $k<k_c$.} 
    In the limit $\gamma\to0^+$ (black curve), $m_2(k)$ has a simple expression given in \eqref{m2_gamma0}. In this case, the function $m_2(k)$ is discontinuous at $k_c$, but it is smooth for any $\gamma>0$. Top right: Same plot zoomed around $k_c$, and for values of $\gamma$ close to $\gamma_c$. While there seems to be a cusp at $k_c$ for small $\gamma>0$ when looking at the large scales, the function $m_2(k)$ appears to be smooth when zooming sufficiently. 
    Bottom left: Plot of $k$ versus $k_0$ {\blue (obtained by the same method)}. When $\gamma$ is smaller than some critical value $\gamma_c= 0.1787369...$, the function $k_0(k)$ becomes non-monotonous close to $k_c$. The inverse function $k(k_0)$ thus becomes multi-valued, leading to the coexistence of 2 stable and one unstable steady states for the same value of $k_0$, corresponding to different values of $k$. The value $\gamma_c=0.1787369...$ was computed numerically, using as a criterion that the minimum of $\frac{dk_0}{dk}$ vs $k$ vanishes at $\gamma_c$ (it is negative for $\gamma<\gamma_c$ and positive for $\gamma>\gamma_c$). For $\gamma=0^+$, the bistability occurs in the interval $k_0\in [6.614585...,9.449407...]$. Bottom right: Same plot zoomed around $k_c$, and for values of $\gamma$ close to $\gamma_c$. The region where the function $k_0(k)$ is non-monotonous seems to be entirely located at $k<k_c$.}
    \label{m2vsk}
\end{figure}

In the above section we have obtained the stationary density $\rho_s(x)$ as
a function of the parameter $k$, with a transition occurring at $k=k_c$ between
a single support for $k<k_c$ and a disjoint support for $k>k_c$. This allowed us
to obtain $k_0$ as a function of $k$, i.e. $k_0= k + 3 m_2(k)$ where
$m_2(k)$ is the second moment of $\rho_s$ as a function of $k$. 
Expressions for $m_2(k)$ in terms of ratios of simple integrals 
where given in Eqs. \eqref{k0versusk1}, \eqref{k0versusk_critical} for $k \geq k_c$,
and a corresponding expression for $k<k_c$ can be obtained from \eqref{rhos_joint} as
explained there.  
It is important to now study how $k$ depends on $k_0$, i.e. to invert
this relation. By numerically computing these integrals 
we find that there are two cases which are illustrated
in Fig.~\ref{m2vsk}:
\begin{itemize}
\item For $\gamma > \gamma_c=0.1787369...$, one finds that $\frac{dk_0}{dk} >0 $, hence
there is always a unique solution for $k$ as a function of $k_0$. 
\item For $\gamma < \gamma_c$, there is a region of values of
$k_0$, corresponding to $k$ close to $k_c$, where there are several solutions for $k$ as a function of $k_0$,
i.e. there are several values of $k$ which correspond to the same
$k_0$, see Fig.~\ref{m2vsk}. This is the bistable regime.
\end{itemize}
The value of $\gamma_c$ was determined numerically as the value of $\gamma$ for which the minimum of $\frac{dk_0}{dk}$ as a function of $k$ (which is a monotonously increasing function of $\gamma$) is exactly zero.

The asymptotic behaviors of $k$ versus $k_0$ can be obtained analytically. Indeed, the behavior of the edges $|y_{1,3}|\sim1/|k|$ as $k\to-\infty$ and $|y_{1,3}|\sim\sqrt{k}$ as $k\to+\infty$ implies $m_2 \sim 1/k^2$ as $k\to-\infty$ and $m_2\sim k$ as $k\to+\infty$, leading to $k\simeq k_0$ for $k\to-\infty$ and $k\simeq k_0/4$ for $k\to+\infty$. However the relation between $k$ and $k_0$ depends on $\gamma$ for $k=O(1)$. In particular, $k$ vanishes for some strictly positive value of $k_0$ (which seems to increase as $\gamma$ decreases).


\subsubsection{Description of the transition in the absence of bistability}

In Sec.~\ref{sec:rho_derivation}, we have seen that there is a transition at $k=k_c=3/2^{2/3}$ between a phase where the density has a joint support $[-y_1,y_1]$ for $k<k_c$, and a phase with disjoint support $[-y_1,y_3]\cup[-y_3,y_1]$ for $k>k_c$, where the $y_i$ are defined in \eqref{yvalues1} for $k>k_c$ and in \eqref{defy1_joint}-\eqref{defy23_joint} for $k<k_c$ (we recall that $y_3<0$). Note that $y_1$ is a continuous function of $k$ even at $k=k_c$ (see Fig.~\ref{phase_diagram_doublewell}), so that the external edges of the support vary continuously. However $y_3$ does not vanish as $k\to k_c$. Instead the interval $[y_3,-y_3]$ fills in continuously when $k$ decreases from $k_c$. We find, from \eqref{rhos_joint}, that the density at the center $\rho_s(0)$ vanishes as $k \to k_c^-$
as
\be \label{rhosx0}
\rho_s(0) \propto \exp \left( -\frac{2^{1/3} \pi \gamma}{\sqrt{3(k_c-k)}} \right) \;,
\ee
where we have used that $y_1\simeq 2^{2/3}(1-\frac{2^{4/3}}{9}(k_c-k))$, leading to $3y_1^2-4k\simeq \frac{4}{3}(k_c-k)$,
and that $K$ is strictly positive at $k=k_c$, hence we can neglect its $k$ dependence at leading order. The density thus exhibits an essential singularity at $k_c$. 
We have plotted $m_2$ as a function of $k$ and as a function of $k_0$ in Fig.~\ref{m2vsk}, and
one can see that these curves appear quite smooth around $k_c$. This is consistent with the fact that the density at zero exhibits only an essential singularity.

Let us however recall that the true physical parameter that we should consider is not $k$, but rather its non-renormalized counterpart $k_0$. For $\gamma>\gamma_c$, the mapping between $k_0$ and $k$ is smooth and single-valued, so that the picture does not change qualitatively when using $k_0$ as a parameter instead of $k$, i.e. in this case the order of the transition is still infinite. The case $\gamma<\gamma_c$ should however be discussed separately, which we now do.

\subsubsection{Bistability} \label{sec:bistability}

\begin{figure}
    \centering
    \includegraphics[width=0.45\linewidth]{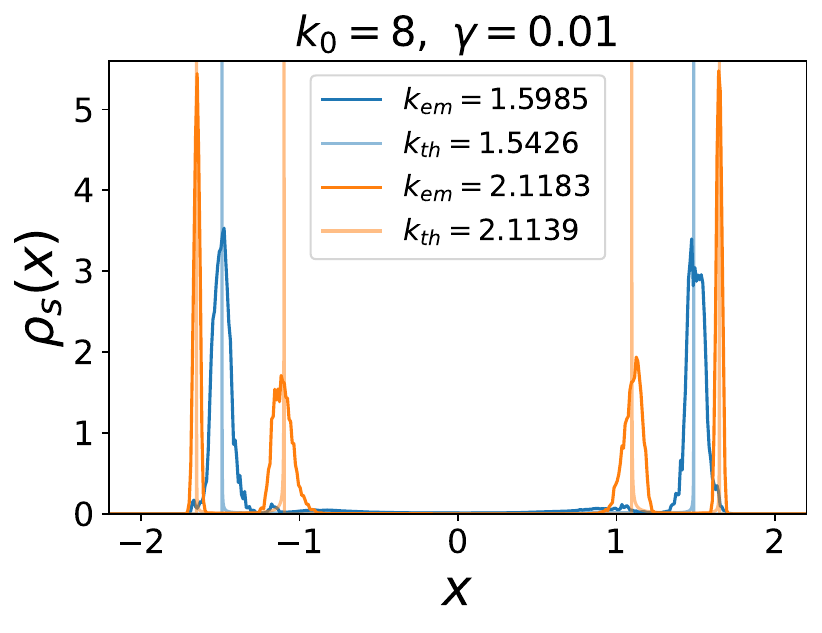}
    \caption{Example where two steady states are observed in simulations for the same values of the parameters ($k_0=8$, $\gamma=0.01$, $g=1$, $v_0=1$ and $N=100$). The two densities were obtained using the same uniform initial condition with support $[-1,1]$, {\blue but for different realizations of the telegraphic noise}. One of steady states (in blue) has a single support, while the other one (in orange)
    has a disjoint support. 
    For each of the two steady states, the measured values of $m_2$ and $k$ ($k_{em}$ in the legend of the figure) coincide quite well with the two possible values predicted theoretically ($k_{\rm th}$ in the
    legend). The lighter lines show the two theoretical predictions for $N\to+\infty$ for the two possible values of $k$,
    which for such a small value of $\gamma$ are very close to delta peaks at $x=\pm y_1$, and at $x= \pm y_3$ in the case of the disjoint support.}
    \label{fig_bistable}
\end{figure}

{\blue For $N\to+\infty$ and $\gamma<\gamma_c$, there exists an interval of values of $k_0$ around $k_{0,c}$ where there are three possible values of $k$\footnote{Note that for a given value of $k$ there is however a unique corresponding value of $k_0$, hence in particular the critical value $k_{0,c}=k_0(k_c)$ is uniquely defined.}, 
see Fig.~\ref{m2vsk} (bottom). There are two possible situations.
(i) For most values of $\gamma$ and $k_0$, 
two of these stationary state have a connected support ($k<k_c$) and the other one has a disconnected support ($k>k_c$). 
(ii) In a small region of parameters $k_0$ and $\gamma$ however, we have noticed 
that all three solutions correspond to $k<k_c$, i.e. to a
connected support, see Fig.~\ref{m2vsk} (bottom right). All three values correspond a priori to possible stationary states of the system, which however can be stable or unstable. We have not studied analytically the stability of these solutions (which seems a priori technically challenging). 
However, we have compared these predictions with numerical simulations for finite $N$ as we now discuss.

The intermediate value, which is always very close to $k_c$, seems to be unstable and was never observed in the simulations. In case (i), the other two values however can both be realized, depending on the initial condition and on the realization of the noise, leading to two possible stationary states. An example from numerical simulations is shown in Fig.~\ref{fig_bistable} (for $N=100$ particles), where either one or the other stationary state is reached depending on the realization, starting from the same initial condition. The case (ii) is more difficult to observe in numerical simulations as it corresponds to a very limited range of parameters, and it is not clear whether both solutions are indeed stable in this case.


Note that an important consequence of this bistability is that in some range of parameters the transition between a connected and disconnected support is now discontinuous as a function of $k_0$.
Finally, there are two interesting questions for future investigations. It would be nice to be able to prove the stability properties of the
different stationary solutions at infinite $N$. It is an open question whether at finite $N$ the system
can transition from one stationary state to the other over an extremely long period of time.} 
\\

{\bf Limit $\gamma \to 0^+$.} It is possible to obtain more quantitative results concerning this bistability in the limit where $\gamma \to 0^+$. In that limit the particles spend most of the time in the
vicinity of the fixed points $\pm y_1$ and $\pm y_3$, and 
the stationary density becomes
\bea \label{gamma0sym}
&& \rho_s(x) = \frac{1}{4}  \left( \delta(x+y_1) + \delta(x-y_3) + 
\delta(x-y_1) + \delta(x+y_3) \right)   \quad , \quad k \geq k_c \;, \\
&& \rho_s(x) = \frac{1}{2}  \left( \delta(x+y_1) +
\delta(x-y_1)  \right)   \quad , \quad k < k_c \;.
\eea 
The fact that all weights are equal is a consequence of our assumption that
the stationary density is even and of the symmetry between $+$ and $-$ particles. Note that for $k=k_c$ the points $\pm y_1=\pm y_2$ are still fixed points of the dynamics, so that this limiting case is in the continuity of $k>k_c$.
One thus has, denoting $y_i=y_i(k)$ to emphasize their dependence in $k$,
\bea \label{m2_gamma0}
&& m_2(k)= \frac{1}{2} (y_1(k)^2 + y_3(k)^2 )  \quad , \quad k \geq k_c \;, \\
&& m_2(k) = y_1(k)^2 \quad , \quad k < k_c \;, \nn
\eea 
The function $m_2(k)$ is right-continuous at $k=k_c=3/2^{2/3}$, but it exhibits a jump at the left, equal to
\be 
\Delta m_2 = m_2(k_c)-m_2(k_c^-) = - \frac{1}{2} (y_1(k_c)^2 - y_3(k_c)^2) = - \frac{k_c}{2} = - \frac{3}{2^{5/3}} \;,
\ee 
since $y_1(k_c)^2=2^{4/3}$ and $y_3(k_c)^2=2^{-2/3}$. This jump becomes smooth for $\gamma>0$, as we have already discussed and as can be seen in Fig.~\ref{m2vsk}. This jump implies that, for some values of $k_0$, there are 2 possible values of $k$
which satisfy
\be 
k_0 = k + 3 m_2(k) \;,
\ee 
one for $k<k_c$ and one for $k\geq k_c$. For $\gamma=0^+$ this happens in the range 
\be k_0 \in [k_c + \frac{3}{2} (y_1(k_c)^2+ y_3(k_c)^2)  , k_c + 3 y_1(k_c)^2  ] = [ \frac{21}{2^{5/3}}, \frac{15}{2^{2/3}}  ] = [6.6145..., 9.4494...] \;.
\ee 

When the value of $\gamma$ is small but not $0^+$, the curve $k_0(k)$ becomes smooth
and in the bistable region a third solution appears near $k=k_c$. This solution
however is presumably unstable. Away from $k_c$ the value of $m_2(k)$ does not change much as long as $\gamma \ll 1$. Around the jump at $k=k_c$ however, for
the value of $m_2(k)$ is very sensitive to a small increase in $\gamma$.
In fact the two limits $\gamma \to 0^+$ and
$k \to k_c$ do not commute, since the relaxation time for the convergence to the fixed
point $\pm y_3$ vanishes at $k_c$. This can be seen for instance
in \eqref{rhosx0}. One can thus expect that $m_2(k)$
will take a crossover form in the double limit $\gamma \to 0$, $k \to k_c$, 
as a function of $\gamma/|k-k_c|^\alpha$
with some unknown exponent $\alpha$. 

As a side remark, we note that in the limit $\gamma\to 0^+$, the values of $y_1$ and $y_3$ can be obtained explicitly as a function of the bare interaction parameter $k_0$ by inserting the relation \eqref{m2_gamma0} into the cubic equation \eqref{cubic_eq} (using that $k=k_0-3m_2$). For $k<k_c$ this gives that $y_1$ is the only real root of
\be \label{eq_y1_gamma0}
4 y_1^3 - k_0 y_1 - 1 = 0 \;,
\ee
while for $k>k_c$, we obtain the coupled equations for $y_1$ and $y_3$
\bea \label{eq_y1y3_gamma0_1}
\frac{5}{2} y_1^3 + \frac{3}{2} y_1 y_3^2 - k_0 y_1 - 1 = 0 \;, \\
\frac{5}{2} y_3^3 + \frac{3}{2} y_1^2 y_3 - k_0 y_3 - 1 = 0 \;. \label{eq_y1y3_gamma0_2}
\eea


%
%
%

\subsubsection{Edge Behavior}

We now consider other general properties of the density, independently of the transition and bistability discussed above. The results of this section and the next are valid in both phases, for any values of $k$ and $\gamma$.

Let us first consider the behavior of the stationary density at the edges, as a function of $k$, in both phases. The considerations below are summarized in Fig.~\ref{phase_diagram_doublewell} of the introduction (blue lines). We have seen in Sec.~\ref{sec:rho_derivation} that a change of behavior always occurs at $\gamma = \gamma_1(k):= 3 y_1^2 -k$. We now show that $\gamma_1(k)$ has a single minimum as a function of $k$, which is strictly positive. To this aim, let us recall that $y_1$ is the largest root of the cubic equation $y_1^3 - k y_1 - 1 = 0$. Taking the derivative with respect to $k$ leads to
\be \label{dy1dk}
\frac{dy_1}{dk} = \frac{y_1}{3y_1^2-k} \;.
\ee
Now, we want to find the solution $k^*$ of the equation $\frac{d}{dk} (3 y_1^2 - k) = 0$, i.e. $6 y_1 \frac{dy_1}{dk} = 1$. Using \eqref{dy1dk}, this becomes $3y_1^2+k=0$. Replacing $k$ in the cubic equation for $y_1$, we obtain $y_1^*=2^{-2/3}$, and thus
\be
k^* = -\frac{3}{2^{4/3}} \quad , \quad \gamma^*=\gamma_1(k^*) = 
\frac{3}{2^{1/3}} \;.
\ee
Note that this computation is valid both for $k<k_c$ and $k>k_c$, so that $k^*$ is the only extremum of $\gamma_1(k)$ on the whole real axis. In addition the asymptotics of $y_1$ are given by 
\be
y_1(k) \sim \begin{cases} \sqrt{k}+\frac{1}{2k} +O(k^{-5/2}) \quad \text{for } k\to+\infty \;, \\ \frac{1}{|k|} + O(k^{-3}) \quad \text{for } k\to-\infty \;, \end{cases} 
\ee
leading to $\gamma_1(k) \sim 2k$ for $k\to+\infty$ and $\gamma_1(k)\sim|k|$ for $k\to-\infty$, thus $k^*$ is indeed a minimum\footnote{Note also that the same method can be used to show that $b=y_1^2-k$ is strictly positive for any $k<k_c$. Indeed, one finds that the solutions of $\frac{d}{dk} (y_1^2 - k) = 0$ should satisfy $y_1^2=k$, which is incompatible with the cubic equation. This proves that $b$ is a monotonously decreasing function of $k$, and thus it is always greater than its value at $k_c$, i.e. $b>2^{-2/3}$.}.

The implication of these results is that, for $\gamma<\gamma^*$, one always has $\gamma/(3y_1^2-k)<1$, and thus the density always diverges at the edges $\pm y_1$. By contrast, for $\gamma>\gamma^*$, there is an interval $[k_1,k_2]$, with $k_1<k^*<0$ and $k_2>k^*$, of values of $k$ such that the density vanishes at $\pm y_1$. The size of this interval increases with $\gamma$. Note that one has $k_2=-\gamma+y_1^2>-\gamma$, which gives a lower bound for $k_2$. 

A natural question is whether a similar threshold value $\gamma^*$ exists for the interior edges $\pm y_3$, when $k>k_c$. The computation performed above for $y_1$ is also valid for $y_3$ since it only uses the fact that it is a root of the cubic equation. This shows that the equation $\frac{d}{dk}(3y_3^2-k)=0$ has no solution on $[k_c,+\infty[$, and thus $\gamma_3(k)=3y_3^2-k$ is a monotonously increasing function of $k$ on this interval. In addition, one has $y_3=-1/2^{1/3}$ for $k=k_c$, thus $\gamma_3(k_c)=0$, and $y_3(k)\sim -\sqrt{k}+\frac{1}{2k} +O(k^{-5/2})$ for $k\to+\infty$, thus $\gamma_3(k) \sim 2k\to+\infty$. Therefore the equation $\gamma=3y_3^2-k$ always has a unique solution $k_3$ for $k>k_c$, meaning that there is always an interval $[0,k_3]$, with $k_3>k_2$, such that the stationary density vanishes at the internal edges $\pm y_3$, even for small $\gamma$ (there is no threshold value analog to $\gamma^*$ in this case). Both curves $\gamma_1(k)$ and $\gamma_3(k)$ are plotted in Fig.~\ref{phase_diagram_doublewell} (right panel). These curves separate the region where the density diverges from the one where it vanishes, at the edges $\pm y_1$ and $\pm y_3$ respectively.



\subsubsection{Convexity around $x=0$}

As we have seen in Section \ref{sec:brownian}, the transition at $k_c$ is absent in the presence of Brownian noise $T>0$, since it is due to the boundedness of the RTP noise. Similarly, the existence of divergences at the edges of the support are due to the persistence of the RTPs and are also absent in the Brownian case. However, as discussed above the passive case also has an interesting change of behavior at $T>0$ between a unimodal and a bimodal distribution, which occurs at $k=0$ (which we recall corresponds to a nontrivial value of $k_0$). In the passive case the equilibrium density is $\rho_{eq}(x) = Ke^{- \frac{1}{T}(-\frac{k}{2} x^2 + g \frac{x^4}{4})}$ (see above), which has a single maximum at $x=0$ for $k\leq 0$, and two maxima at $x=\pm \sqrt{k/g}$ for $k>0$ (with a local minimum at $x=0$). It is therefore interesting to study the convexity of $\rho_s(x)$ around $x=0$ to see if such a change of behavior also exists in the active case. Of course, this question is only relevant for $k<k_c$, where there is a single support.

Here we simply need to compute the second derivative of $\ln \rho_s(x)$ at $x=0$ using \eqref{rhos_joint}. We find
\be 
\partial_x^2 \ln \rho_s(x)|_{x=0} 
= \frac{k \left(k +\gamma\right)}{ y_1^2(y_1^2-k)^2} \;.
\ee 
As in the passive case, this expression is positive for $k>0$, indicating that $x=0$ is a local minimum of the density, and negative for $-\gamma<k<0$, indicating that $x=0$ is a local maximum (the denominator is always strictly positive). However, here there is an additional change of behavior, where $x=0$ becomes again a local minimum for the density for $k<-\gamma$. This additional crossover is due to the accumulation of particles at the edges of the support for large persistence time.
Note that the transition at $k=0$ corresponds to a nontrivial value of $k_0$, which can be computed using \eqref{rhos_joint} with $y_1=1$, i.e.
\be
\rho_s(x) = K (1-x^2)^{\frac{\gamma}{3}-1} \left((x^2+1)^2 - x^2 \right)^{-\frac{\gamma}{6}-1} \exp \left( \frac{\gamma}{\sqrt{3}} \left( \arctan \left(\frac{2 x-1}{\sqrt{3}}\right) + \arctan \left( -\frac{2x+1}{\sqrt{3}} \right) \right) \right)
\ee
from which one can compute the corresponding value of $k_{0}=3m_2$.





\section{Non symmetric steady state} \label{sec:asym}

Until now we have assumed that the stationary density $\rho_s(x)$ is symmetric, $\rho_s(-x)=\rho_s(x)$. This should be true in the large $N$ limit if we start from a symmetric initial condition. {\blue It should also hold whenever the support has a single component, due to ergodicity.} However, as we now show, 
if the initial condition is not symmetric, and if the support is disjoint, a non-symmetric stationary distribution is possible, {\blue where the particles are separated into two groups of different sizes}. Such a distribution is characterized by a non-zero value of the third moment $m_3$. 

\subsection{General properties} 

Let us search for a non-parity invariant solution of the self-consistent equation for $\rho_s(x)$ \eqref{self_intro}.
We still impose $m_1=0$ by choosing the origin to be the center of mass,
but one can now have a non zero $m_3$. It turns out that the parameter $m_3$
can take a continuous range of values. To understand why we go back to
the self-consistency condition which determine the stationary density, 
which we write as follows 
\bea
&&2 \gamma F(x) \rho_s(x)  = \partial_x ( (v_0^2- F(x)^2) \rho_s(x)) \;, \label{consisteqm3_1} \\
&&F(x) = \tilde F(x) := - \int dx' W'(x-x') \rho_s(x') \;, \label{consisteqm3_2}
\eea 
where
\bea \label{Ftilde_m3}
 && \tilde F(x)  = -  \int dy (  - k_0 (x-y) + (x-y)^3 ) \rho_s(y)   = m_3 + (k_0-3 m_2) x  - x^3 \;,
\eea  
{\blue is the force that is obtained from the steady state density.}
The idea is to solve the first equation \eqref{consisteqm3_1} for $\rho_s(x)$ with some arbitrary given $F(x)$, and then
insert the solution in the r.h.s. of the second equation \eqref{consisteqm3_2}, which leads to
a self consistent condition for $F(x)$, hence for $\rho_s(x)$.
In practice we see that we can choose $F(x)=\mu_3 + k x - x^3$,
where $k$ and $\mu_3$ must be determined by \eqref{consisteqm3_2}. Equation \eqref{Ftilde_m3} then gives the condition $k=k_0-3m_2$, as in the symmetric case, as well as $\mu_3=m_3$, where $m_2$ and $m_3$ should be interpreted as functions of $k$ and $\mu_3$.

However, even before imposing this self-consistency condition \eqref{consisteqm3_2}, we see that integrating the first equation \eqref{consisteqm3_1} over $\mathbb{R}$ one
obtains, using that $\rho_s(x)$ vanishes at infinity,
\be 
\int dx F(x) \rho_s(x) = 0 \;,
\ee
which, from \eqref{rhod}, is equivalent to $\int dx \rho_d(x)=0$, i.e. equilibration
between the two species $\pm v_0$.
Using that
\be
\int dx F(x) \rho_s(x) = \int dx (\mu_3 + k x - x^3) \rho_s(x) 
= \mu_3 +k m_1 - m_3 \;,
\ee
and the fact that we were free to choose $m_1=0$, we see that the equation $\mu_3=m_3$ is always satisfied.
Hence $m_3$ is undetermined and can take a continuous range of values (i.e. any value leading to a disconnected support). 


Let us recall that we are considering here the case where the support of the density is disjoint.
Let us denote $\Omega_L$ and $\Omega_R$ the two connected components (on the
negative and positive side respectively). Then, one has furthermore 
that $\int_{\Omega^{R/L}} dx \tilde F(x) \rho_s(x)=0$ for each component,
which is equivalent to $\int_{\Omega^{R/L}} dx  \rho_d(x)=0$ in each component (i.e. each component equilibrates
separately). We now define the left/right moments 
\be \label{defmnRL}
m_n^{R/L} = \int_{\Omega^{R/L}} dx x^n \rho_s(x) 
\ee 
associated to each component. Using that $\tilde F(x)=m_3+ k x - x^3$ 
and integrating w.r.t. $\rho_s(x) dx$ we obtain the relations
\bea \label{eqmRmL} 
&& m_3 m_0^L + k m_1^L = m_3^L \;, \\
&& m_3 m_0^R + k m_1^R = m_3^R \;, \nn
\eea 
where $m_0^{R/L}$ are the fractions of particles in $\Omega_{R/L}$
with $m_0^R+m_0^L=1$. 

\subsection{Detailed solution and support of the density}

\begin{figure}
\centering
\includegraphics[width=0.5\linewidth,trim={2cm 4cm 1.95cm 4cm},clip]{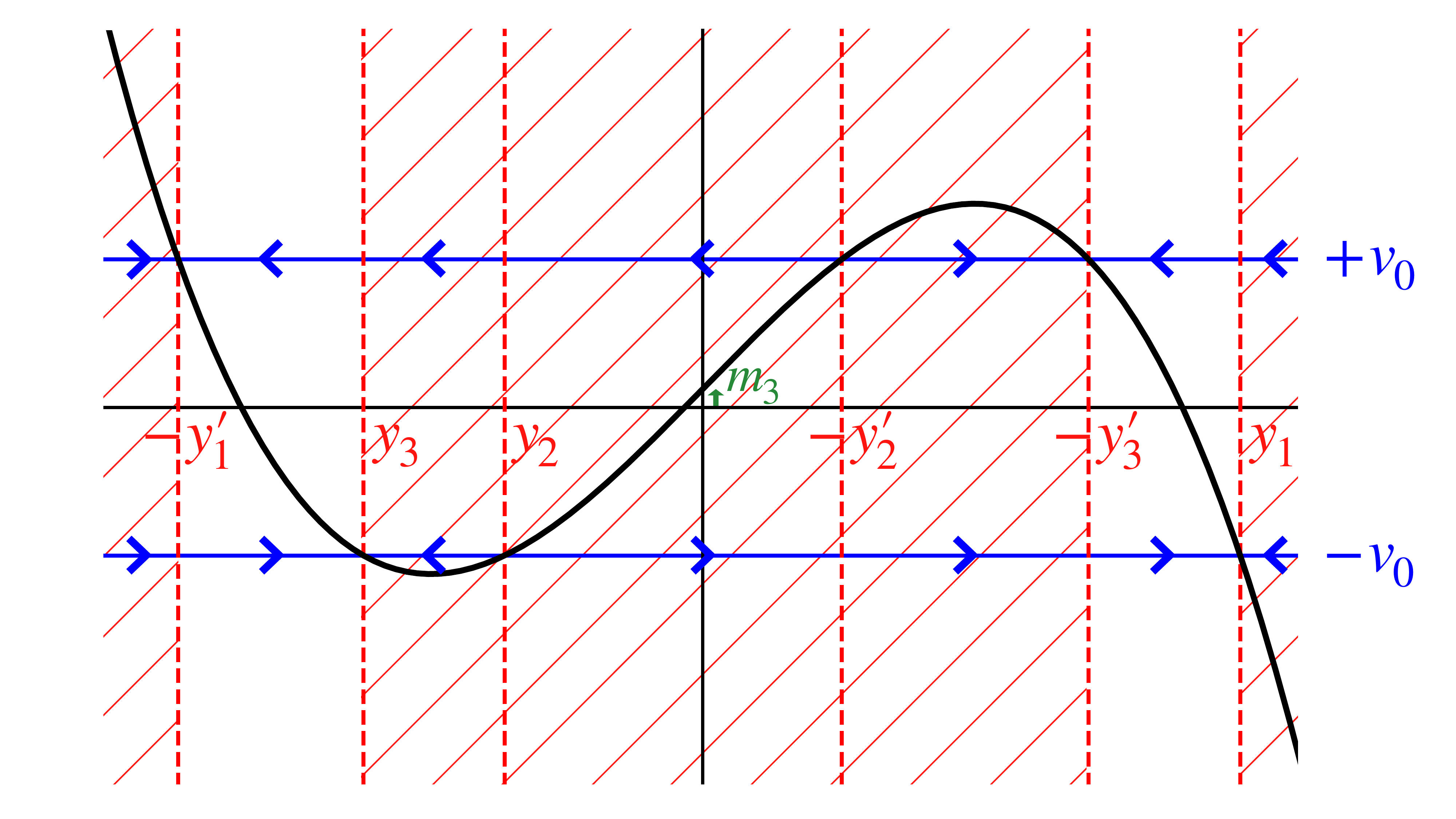}
\caption{Plot of $\tilde F(x)$ (black curve) with $m_3>0$, in the regime $\Delta>0$. The regions hatched in red are inaccessible to the particles in the stationary state. Compared to the case $m_3=0$ (central panel of Fig.~\ref{stability_diagram}), the presence of a strictly positive $m_3$ corresponds to an upwards shift of the curve $\tilde F(x)$, which creates an asymmetry between the left and right components of the support $[-y_1',y_3]$ and $[-y_3',y_1]$. 
}
\label{stability_diagram_m3}
\end{figure}

We now give the exact solution for the stationary density $\rho_s(x)$, as a function of 
the two parameters $k= k_0-3 m_2$ and $m_3$. 
The support of $\rho_s(x)$ is now determined by the two sets of roots of the two cubic equations
\bea \label{cubic_eqm3}
&& 0 = -1 - \tilde F(y) = y^3 - k y - 1 -m_3 = (y-y_1)(y-y_2)(y-y_3) \;, \\
&& 0 = 1 - \tilde F(y) = y^3 - k y + 1 - m_3 = (y+y_1')(y+y_2')(y+y_3') \;.\label{cubic_eqm3prime}
\eea 
As mentioned above, an asymmetric solution is only possible if the support is disconnected. This already implies that $k>k_c$. In addition, for a given value of $k>k_c$, only a finite range of values of $m_3$ are possible. Indeed, the solution is valid, only if the discriminants of both cubic equations \eqref{cubic_eqm3} and \eqref{cubic_eqm3prime} are positive, i.e. $\Delta = 4k^3 - 27(1\pm m_3)^2 >0$. This gives
\be
|m_3|<m_3^c=\left(\frac{k}{k_c}\right)^{3/2}-1 \quad , \quad k_c = \frac{3}{2^{2/3}} \;.
\ee
Any value of $m_3$ satisfying this condition is a priori possible. In this case, there are two sets of 3 real roots, which are given by
\bea \label{expry1}
y_{n+1} = 2\sqrt{\frac{k}{3}} \cos \left( \frac{1}{3} \arccos \left((1+m_3)\left(\frac{k_c}{k}\right)^{3/2}\right) - \frac{2\pi n}{3}\right) \quad , \quad n=0,1,2 \;, \\
y'_{n+1} = 2\sqrt{\frac{k}{3}} \cos \left( \frac{1}{3} \arccos \left((1-m_3)\left(\frac{k_c}{k}\right)^{3/2}\right) - \frac{2\pi n}{3}\right) \quad , \quad n=0,1,2 \;. \label{expry2}
\eea
These roots are ordered as follows
\be
y_1 > 0 > y_2 > y_3 \quad , \quad y'_1 > 0 > y'_2 > y'_3 \quad , \quad y_1 > -y_3  > - y_2  \quad , \quad y'_1 > - y'_3 > - y'_2 \;.
\ee 
The support $\Omega$ is thus 
\be \label{supportm3} 
\Omega = \Omega_L \cup \Omega_R \quad , \quad \Omega_L = [-y_1',y_3] 
\quad , \quad 
 \Omega_R =  [-y_3',y_1] \;,
\ee
see Fig.~\ref{stability_diagram_m3}. 

As in the symmetric case, the formal solution for \eqref{consisteqm3_1} reads
\be \label{rhos_formal_asym}
\rho_s(x) = \frac{K}{(1-\tilde F(x))(1+\tilde F(x))}
\exp \left( \gamma \int^x dy \frac{1}{1-\tilde F(y)} - \gamma \int^x dy \frac{1}{1+\tilde F(y)} \right) \;, 
\ee 
where
\bea 
&& \int^x dy \frac{1}{1-\tilde F(y)} -  \int^x dy \frac{1}{1+\tilde F(y)} \\
&& = \frac{1}{(y'_1-y'_2)(y'_1-y'_3) } \log |x - y'_1| + \text{2 perm} 
+ \frac{1}{(y_1-y_2)(y_1-y_3) } \log |x + y_1| + \text{2 perm} \;.
\eea 
Hence one now defines two sets of exponents
\bea  
&& \eta_1= \frac{\gamma}{(y_1-y_2) (y_1-y_3) } - 1 = \frac{\gamma}{3 y_1^2-k} - 1 \quad  \text{and 2 perm} \;, \\
&& \eta_1'= \frac{\gamma}{(y_1'-y_2') (y_1'-y_3') } -  1 = \frac{\gamma}{3 (y'_1)^2-k} - 1 \quad  \text{and 2 perm} \;.
\eea  
Here we have used that $-\tilde F'(y)=(y-y_1)(y-y_2)+(y-y_1)(y-y_3)+(y-y_2)(y-y_3) = 3y^2-k$ to rewrite the exponents $\eta_i$ and $\eta_i'$. Putting everything together, one can write 
\be \label{rhos_asym1}
\rho_s(x) = K_R \tilde \rho^R_s(x) + K_L \tilde \rho^L_s(x) \;,
\ee 
where 
\bea \label{rhos_asym2}
\tilde \rho_s^R(x) &=&  (y_1-x)^{\eta_1} (x+y'_3)^{\eta'_3} (x-y_2)^{\eta_2} (x-y_3)^{\eta_3} (x+y'_1)^{\eta'_1} (x+y'_2)^{\eta'_2}  \theta(-y'_3<x<y_1) \;, \\
\tilde \rho_s^L(x) &=&  (y_3-x)^{\eta_3} (x+y'_1)^{\eta'_1} (y_1-x)^{\eta_1} (y_2-x)^{\eta_2} (-y'_2-x)^{\eta'_2} (-y'_3-x)^{\eta'_3}  \theta(-y'_1<x<y_3) \;, \nn
\eea
and $K_R$ and $K_L$ are two normalisation constants. They
are determined from the two conditions
\bea  \label{eqKRKL}
&& K_R \tilde m_0^R + K_L \tilde m_0^L = 1 \;, \\
&& K_R \tilde m_1^R + K_L \tilde m_1^L = 0 \;, \nn
\eea  
where we have defined $\tilde m_n^{R/L} = \int dx x^n \tilde \rho_s^{R/L}(x)$
(note that $m_n^{R/L} = K_{R/L} \tilde m_n^{R/L} $). Hence for
each value of $(k,m_3)$ all the constants are determined.

As in the symmetric case, one can determine a posteriori the corresponding value of $k_0=k+3m_2$. Note that $m_2$ now depends on $m_3$, so that the mapping between $k$ and $k_0$ is different for each value of $m_3$ (and in particular it differs from the symmetric case $m_3=0$). By contrast, the second ``effective'' parameter $m_3$, which controls the asymmetry of the solution, cannot be mapped directly to one of the ``true" parameters of the model. One can compute a posteriori the proportion $m_0^R$ of particles on the right side of the support, which is another, perhaps more intuitive way to quantify the asymmetry of the steady state. 
However, this asymmetry is determined in a non-trivial way by the initial condition, as well as by the history of the noise in the case of large but finite $N$ that we consider in the simulations (see below). The detailed study of this dependence goes beyond the scope of this paper.

\subsection{Limit $\gamma \to 0^+$} \label{subsec:gamma0asym}

\begin{figure}
    \centering
    \includegraphics[width=0.45\linewidth]{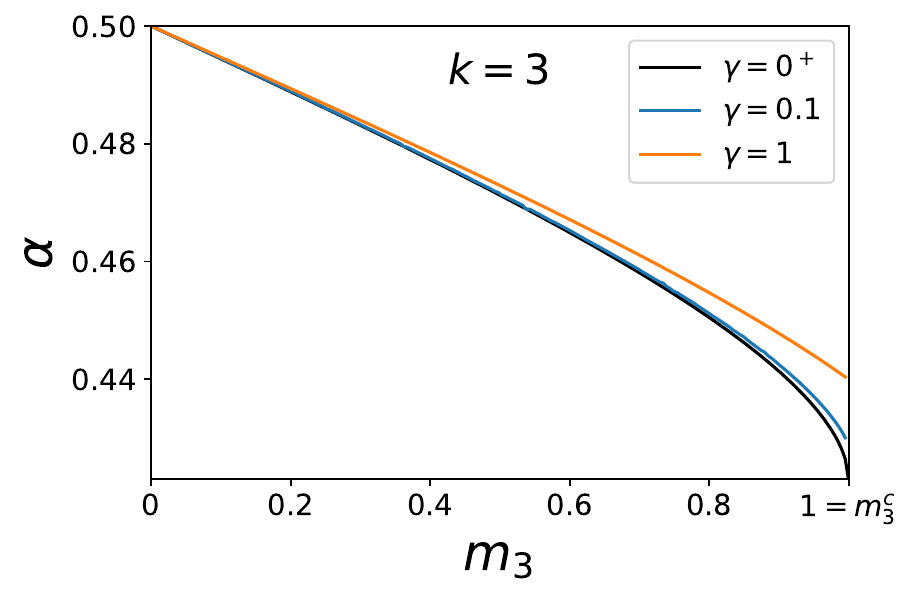}
    \includegraphics[width=0.475\linewidth]{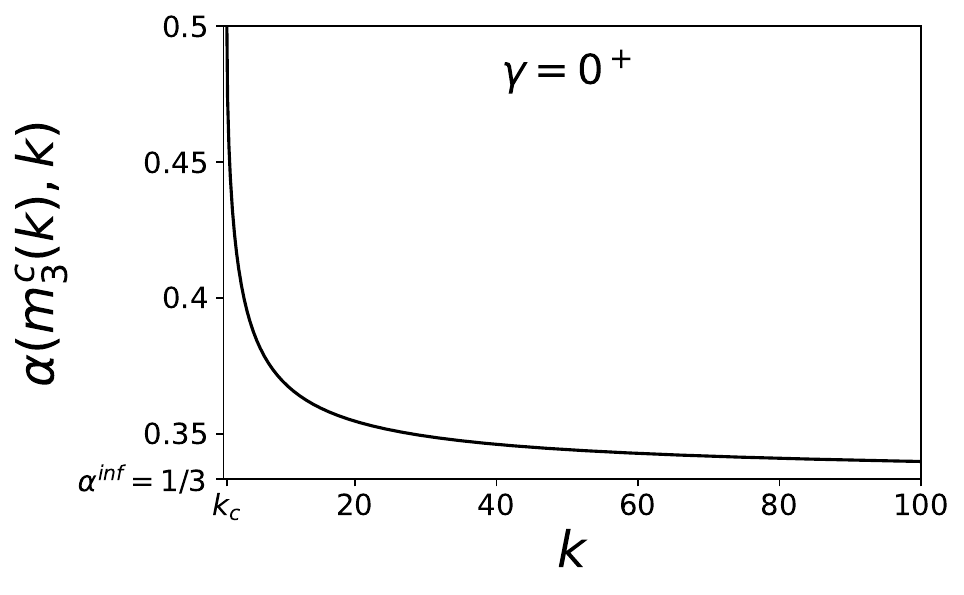}
    \caption{Left: Proportion $\alpha=m_0^R$ of particles on the right component of the support as a function of $m_3$, for $k=3$ (other values of $k>k_c$ give qualitatively similar behaviors). The curve corresponding to the limit $\gamma\to0^+$ was obtained from the expression \eqref{alpha_expr}, while the ones corresponding to $\gamma>0$ were computed numerically using the expression of $\rho_s(x)$ given in \eqref{rhos_asym1}-\eqref{rhos_asym2}. We find that (i) the limit $\gamma\to 0^+$ gives a lower bound for the value of $\alpha$ and (ii) $\alpha$ is always a monotonously decreasing function of $m_3$. Right: Minimal value of $\alpha$, reached for $m_3=m_3^c$ and in the limit $\gamma\to0^+$, as a function of $k$. It decreases with $k$ towards the limiting value $\alpha^{inf}=1/3$, which gives the general lower bound for $\alpha$.}
    \label{fig_alpha}
\end{figure}

In the limit $\gamma \to 0^+$ one can perform more explicit calculations. 
In that limit, 
the system spends all the time near the fixed
points and the stationary density becomes
\be 
\rho_s(x) = \frac{1- \alpha}{2} \left( \delta(x+y'_1) + \delta(x-y_3)  \right) 
+ \frac{\alpha}{2} \left( \delta(x+y'_3) + \delta(x-y_1) \right) \,,
\ee
where $\alpha=m_0^R$ is the total weight (fraction of particles) in the positive part of the support \eqref{supportm3}, which we compute below. Contrary to \eqref{gamma0sym}, there is now an asymmetry between the left and right connected components of the support (parametrized by $\alpha$). However, the symmetry between $+$ and $-$ particles still imposes that the weights of the delta peaks at $y_1$ and $-y_3'$ (resp. $-y_1'$ and $y_3$) are equal.
Hence we have
\bea 
&& m_1^R = \frac{\alpha}{2} (y_1-y'_3) \quad , \quad m_1^L = \frac{1-\alpha}{2} (y_3-y'_1) \;, \label{m3RL1} \\
&& m_3^R = \frac{\alpha}{2} (y_1^3-(y'_3)^3) \quad , \quad m_3^L = \frac{1-\alpha}{2} (y_3^3-(y'_1)^3) \;,
\label{m3RL2}
\eea 
where the $m_n^{R/L}$ are defined in \eqref{defmnRL}. 

Using the cubic equations \eqref{cubic_eqm3} for $y=y_{1/3}$ and 
\eqref{cubic_eqm3prime} for $y=-y'_{1/3}$, the equations \eqref{m3RL1}-\eqref{m3RL2} yield 
\bea 
&& m_3^R = k m_1^R + \alpha m_3 \;, \\
&& m_3^L = k m_1^L + (1- \alpha) m_3 \;.
\eea 
We thus recover the equations \eqref{eqmRmL}, with $m_0^R=\alpha$ and $m_0^L=1-\alpha$, which we have shown above hold for arbitrary $\gamma$.
The condition $m_1=m_1^R + m_1^L = 0$ allows
to determine $\alpha$ from \eqref{m3RL1}, which reads 
\bea \label{alpha_expr}
\alpha = \frac{y'_1-y_3}{y_1'-y_3 + y_1-y'_3} \;.
\eea 
Thus this gives $\alpha=m_0^R$ as a function of the $y_i$ and $y'_i$, i.e. as 
a function of the two parameters $k,m_3$. From \eqref{expry1}-\eqref{expry2}, one can check that for $m_3>0$, one has $y_1>y_1'$ and $-y_3'>-y_3$ (see also Fig.~\ref{stability_diagram_m3}). This implies that, for $m_3>0$, one has $\alpha < 1/2$, i.e. there are more particles on the left than on the right, but that the particles on the right side are further away from $x=0$. Numerically we find that these observations remain true for any $\gamma$ and that $\alpha=\alpha(m_3,k)$ is a decreasing function of $m_3$ for any $k$ (see Fig.~\ref{fig_alpha} -- left panel).


Let us recall that the upper limiting allowed value for $m_3>0$ is
$m_3^c =(k/k_c)^{3/2}-1$. We find that the corresponding value $\alpha(m_3^c,k)$ is a decreasing function of $k$ (see Fig.~\ref{fig_alpha} -- right panel).
For this value we find 
\bea  
&& y_1= 2 \sqrt{\frac{k}{3}} \quad , \quad y_2=y_3= -\sqrt{\frac{k}{3}} \quad , \quad \text{exact} \\
&& y'_1=y'_2= \sqrt{\frac{k}{3}} + o(1)   \quad , \quad y'_3 = - 2 \sqrt{\frac{k}{3}} + o(1) 
\quad , \quad k \to +\infty
\eea  
where the second line is valid for large $k$. 
This implies that $\alpha$ varies in the range
\be
1/3 < \alpha(m_3,k) <  2/3 \;,
\ee 
where the limiting values are given by $\alpha(\pm m_3^c,k\to+\infty)$.

Finally, the observations above can give us an idea of what happens if one starts from an initial condition where $|m_3|>m_3^c$. Indeed in this case, the particles from the side with more particles can access the other side but cannot come back, and thus we expect the number of particles on each side to equilibrate and the value of $|m_3|$ to decrease until it reaches the critical value.




\subsection{Numerical results} 

\begin{figure}
    \centering
    \includegraphics[width=0.45\linewidth]{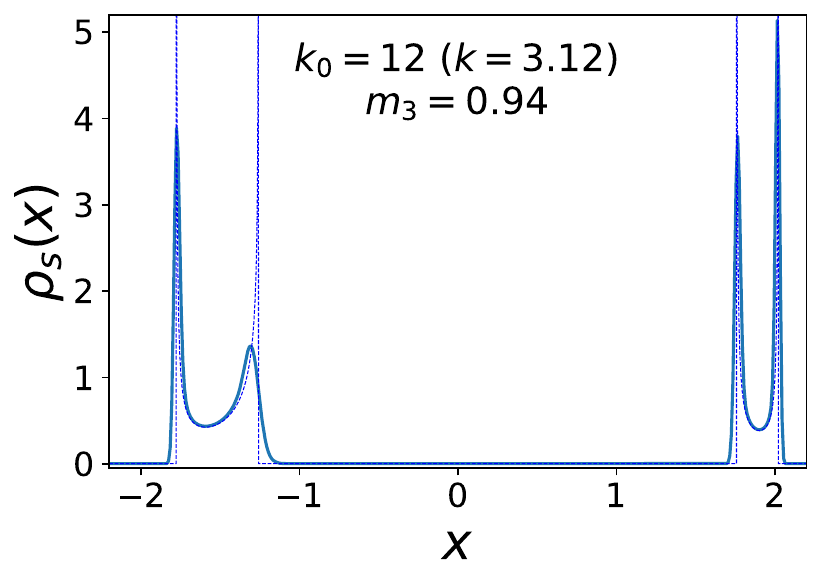}
    \caption{Example of an asymmetric steady state density $\rho_s(x)$, for $k_0=12$, $g=1$, $v_0=1$ and $\gamma=1$. The full line is obtained from a numerical simulation with $N=100$ particles, using an asymmetric initial condition: the particles are initially split into two groups, with 60 particles on the left and 40 on the right, equally separated on the intervals $[-10,-4]$ and $[6,10]$ respectively (the $x$-axis is then shifted such that the center of mass is at $x=0$). From the values of $m_2$ and $m_3$ obtained from the simulation, we also plot the theoretical prediction for $N\to+\infty$ \eqref{rhos_asym1}-\eqref{rhos_asym2} for the corresponding values of $k$ and $m_3$, in dashed lines. We find that the fraction of particles on the right in the stationary state is $\alpha=0.45$ from the simulation and $\alpha=0.445505...$ from the analytical expression, which match up to rounding error $1/(2N)$.}
    \label{fig_asym}
\end{figure}

We have tested some of the predictions above in numerical simulations (for $N=100$ particles). 
One finds that indeed starting from an asymmetric initial condition, e.g. separating the particles into two groups of unequal size at time $t=0$, one can obtain a non-symmetric stationary density, which is well described by the expression \eqref{rhos_asym1}-\eqref{rhos_asym2} --
see Fig. \ref{fig_asym} for an example. In addition we have studied numerically small values of $N$
in Appendix \ref{app:N2}. Interestingly, non-symmetric stationary densities can already be observed for $N=5$. However, they did not arise in our simulations for $N=3$ or $4$, which could be interpreted as a sign that the limiting value $\alpha=1/3$ computed above for $N\to+\infty$ also holds at finite $N$.

\section{Conclusion}

In this paper we have considered $N$ run-and-tumble particles 
in one dimension interacting via a pairwise double well potential $W(r)$,
with associated parameter $k_0=- W''(0)$. 
Thanks to a self consistent equation valid in the large $N$ limit,
we were able to express the stationary density $\rho_s(x)$ in the large $N$ limit as a function of a renormalized parameter $k=k(k_0)$, and of the tumbling rate $\gamma$.
We showed that $\rho_s(x)$ exhibits a transition, where the particles separate into two groups for $k>k_c$. In both phases the density displays power law singularities at the edges of
the support. When the two groups are of the same size, we find that in some range of parameters
around the transition, $k(k_0)$ is multivalued, which implies the 
existence of two stable steady states. Furthermore,
we show that for $k>k_c$ it is possible to observe non-symmetric (non parity invariant)
steady states which are characterized by a fraction $\alpha \neq 1/2$ of
particles in the rightmost group. We show that $\alpha$ can vary continuously inside an interval that we determine. All these analytical predictions, valid for $N \to +\infty$, have been
compared with numerical simulations for $N=100$, with very good agreement.
In particular we have checked that different steady states can be reached
by changing the initial condition. 

It is quite interesting that this simple many body system exhibits a non-uniqueness 
of the stationary state in one space dimension. For the case of thermal Brownian particles 
there is a unique 
Gibbs state at any finite temperature $T>0$, while symmetry breaking
and non-uniqueness only occur at $T=0$. The much richer phenomena observed here seem
to be a consequence of the boundedness of the telegraphic RTP noise. {\blue In particular, we expect that adding a small thermal noise to the model would lead to an unbounded support, 
and transform the non-symmetric steady-states into long lived transients towards a symmetric steady state. It is however less clear what would happen to the bistability discussed in Sec.~\ref{sec:bistability}. Understanding more generally the conditions on the noise and the interaction to observe such non-uniqueness of the stationary state is an open direction for future research.
It would be particularly} interesting to investigate whether such phenomena
occur for similar interactions with other kinds of active particles
and in higher space dimensions. 

Another consideration is that the simple double well interaction considered
here has the unrealistic feature that the interaction force diverges at
infinity. An interesting question is whether one could consider
a more realistic interaction potential such that the effective force
$\tilde F(x)$ has a positive maximum followed by a negative minimum for $x>0$, but vanishes at infinity. We briefly explore this question in Appendix \ref{app:realistic} through numerical simulations. 
Our observations suggest that the phenomenology obtained
here is relevant for that situation as well.

There are several open problems even for this simple type of model. 
For instance one would like to be able to solve for the dynamics
and obtain the basin of attraction of each steady state. Another question is whether, for finite $N$, the noise can allow for rare transitions between these steady states, and at which rate.
One could also try to extend this to higher order degree
polynomials (or more general functions) and determine the co-dimension of the
stationary manifold. 

It would be interesting to investigate further the behavior of
active particles interacting with both repulsive
and attractive components and what kind of structure they can form. In the present case
we found that they can form bound states, which can split into two groups of particles,
which we could call an "active super-molecule".
\\

{\bf Acknowledgments.} We acknowledge support from ANR Grant No. ANR- 23-CE30-0020-01 EDIPS. We thank Naftali Smith for his interesting comments.

\appendix

\section{Self-consistent equation for the stationary density} \label{app:self}

Let us recall briefly the derivation of the self-consistent equation \eqref{self_intro} for the stationary density in the large $N$
limit in the case of the RTPs \cite{activeRDshort}, in the absence of passive noise, i.e. for $T=0$ but with an additional external potential $V(x)$. We first consider a general interaction potential $W(x)$, with $W'(0)=0$, such that the particles can cross. 
In this case, one can use the Dean-Kawasaki method \cite{Dean,Kawa} extended to the RTPs \cite{TouzoDBM2023} to obtain the evolution equation for the densities $\rho_\sigma$ with $\sigma=\pm 1$ in the limit of large $N$,
\be \label{eqrho1n}
 \partial_t \rho_\sigma(x,t)  =  
\partial_x \left[\rho_\sigma(x,t)  \left( - v_0 \sigma +  V'(x) +
\int dy \, W'(x-y) (\rho_+(y,t) + \rho_-(y,t) ) \right) \right]  + \gamma \rho_{- \sigma}(x,t) - \gamma \rho_{\sigma}(x,t) \;.
\ee
In terms of the densities $\rho_s= \rho_+ + \rho_-$ and $\rho_d= \rho_+ - \rho_-$ this leads to the system of equations
\bea \label{eqrho2n}
&& \partial_t \rho_s =    \partial_x [ (- v_0 \rho_d - \tilde F(x,t) \rho_s ] \;, \\
&& \partial_t \rho_d =   \partial_x [ (- v_0 \rho_s - \tilde F(x,t) \rho_d ] - 2 \gamma \rho_d \;, \nn
\eea 
where
\be 
\tilde F(x,t) = - V'(x) - \int dy \, W'(x-y) \rho_s(y,t) \;.
\ee 
The equations \eqref{eqrho2n} are identical to the equations of a single RTP, $N=1$, in a effective time dependent force field
$\tilde F(x,t)$ which depends itself on the time dependent density. One can thus use the known results for this simpler problem,
whenever they are available. 

Let us assume from now on that 
the interaction is sufficiently attractive at large distance so that the 
steady state densities exist and vanish at infinity. Then they are solution of
\bea \label{eqrho22n}
&& 0 =    \partial_x [ (- v_0 \rho_d - \tilde F(x) \rho_s ] \;, \\
&& 0 =   \partial_x [ (- v_0 \rho_s - \tilde F(x) \rho_d ] - 2 \gamma \rho_d \;, \nn
\eea 
where the effective force field $\tilde F(x)$ is now static 
and depends itself on the steady state density. 

These equations can be solved formally since the stationary measure for a single RTP in an arbitrary 
force field is known. Let us recall the main steps. Integrating
the first equation on the real line, assuming that $\tilde F(x) \rho_s(x)$ vanish at
infinity one finds 
\be 
\rho_d(x) = -\frac{\tilde F(x)}{v_0} \rho_s(x) \;. \label{rhodself}
\ee 
Inserting in the first equation we obtain
\be \label{eqdiffrhos} 
2 \gamma \tilde F(x) \rho_s(x)  = \partial_x ( (v_0^2-\tilde F(x)^2) \rho_s(x)) \;.
\ee 
Denoting $\rho_s(x)=f(x)/(v_0^2-\tilde F(x)^2)$, one has $f'(x)=2 \gamma \tilde F(x)/(v_0^2-\tilde F(x)^2) f(x)$.
Integrating this equation we obtain the self-consistent equation 
\be 
\rho_s(x) = \frac{K}{v_0^2-\tilde F(x)^2} e^{2 \gamma \int^x dz \, \frac{\tilde F(z)}{v_0^2-\tilde F(z)^2}}
\quad , \quad 
\tilde F(z)= -V'(x) - \int dy \, W'(z-y) \rho_s(y) \;,
\label{self}
\ee 
where $K$ is a constant determined by normalization.
Here we have assumed that the support of the density is a single interval (which can be infinite). However, depending of the number of roots of the equations $\tilde F(x)=\pm v_0$, the support can be a union of disjoint intervals and one must determine the value of $K$ for each of these intervals, reflecting the fraction of particles that they contain. 
These values may depend on the initial condition. 

Once $\rho_s(x)$ is known, $\rho_d(x)$ is obtained from \eqref{rhodself}. 
From \eqref{rhodself}, since $-\rho_s \leq \rho_d \leq \rho_s$, we see that the support of $\rho_s(x)$
must be included in the region where $|\tilde F(x)| \leq v_0$.
\\



\section{Double-well interaction for $N=2$, and numerics for small values of $N$}
\label{app:N2}

\begin{figure}
    \centering
    \includegraphics[width=0.45\linewidth]{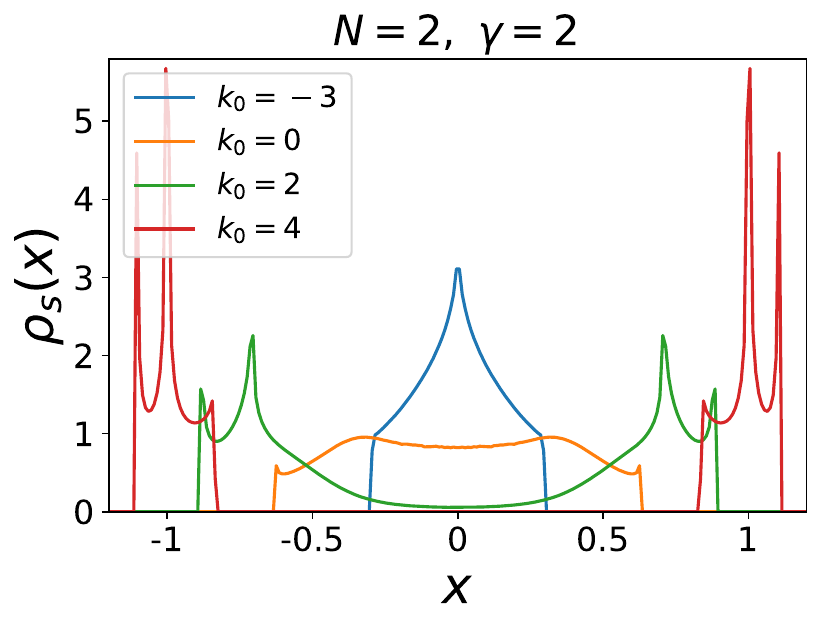}
    \includegraphics[width=0.45\linewidth]{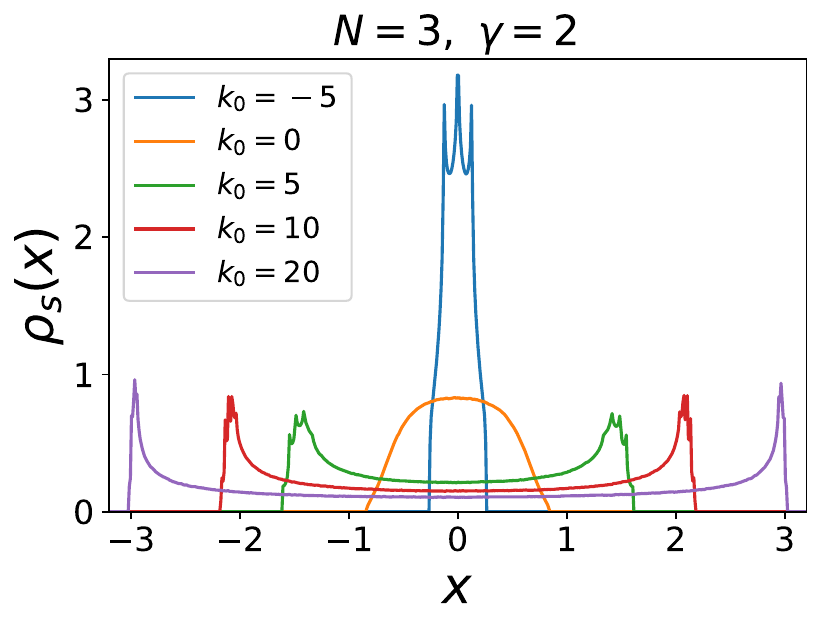}
    \includegraphics[width=0.45\linewidth]{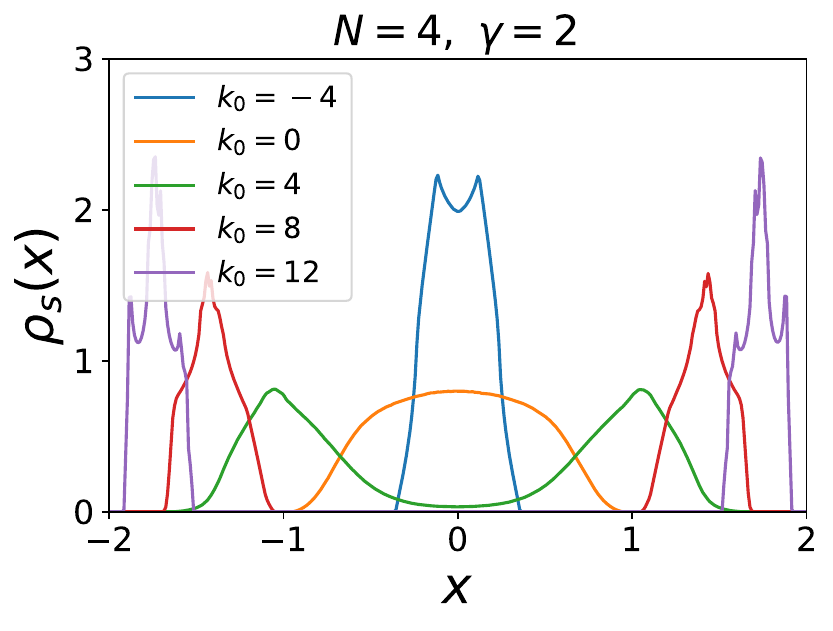}
    \includegraphics[width=0.45\linewidth]{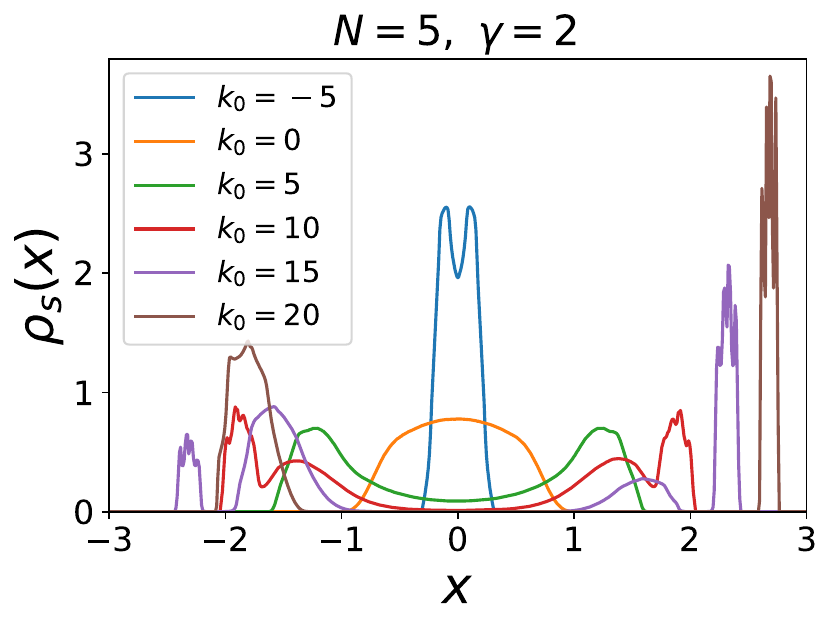}
    \caption{Plots of the stationary density $\rho_s(x)$ for $N=2$, 3, 4 and 5, for $\gamma=2$, $v_0=1$, $g=1$ and different values of $k_0$ (in the frame of the center of mass).
    For $N=3$ we observe that the support is always joint. For $N=2$ and $4$ there is a transition from a joint to a disjoint support as $k_0$ increases, but the density remains symmetric. For $N=5$, there is a spontaneous breaking of the parity symmetry as $k_0$ increases, and the support becomes disjoint with 2 particles on one side and 3 on the other side.}
    \label{fig:smallN}
\end{figure}

In this Appendix we consider 2 RTPs interacting via the double-well potential $W(x)=-\frac{k_0}{2} x^2+ \frac{g}{4} x^4$,
\bea
\dot x_1 = k_0(x_1-x_2) - g(x_1-x_2)^3 + v_0 \sigma_1(t) \\
\dot x_2 = -k_0(x_1-x_2) + g(x_1-x_2)^3 + v_0 \sigma_2(t)
\eea
We 
focus on the difference $X=x_2-x_1$, which evolves according to
\be
\dot X = f(X)  + v_0 (\sigma_2(t) - \sigma_1(t))
\quad , \quad f(X) = 2 k_0 X - 2 g X^3  \;.
\ee

The support of $X$ can be obtained by studying the fixed points of this equation of motion. This is very similar to the case $N\to+\infty$ (see Fig.~\ref{stability_diagram}), but with a non-renormalized $k_0$. In addition to the two lines $f(X)=\pm2v_0$, there are additional fixed points 
corresponding to $f(X)=0$ (corresponding to the $\sigma_1=\sigma_2$), but they do not play any role in determining the support. Thus there is a transition for some values of $k_0$ between a regime where $X$ can take all values inside an interval $[-X_e, X_e]$, meaning that the two particles can cross, and a state where they cannot cross, i.e. the support of $X$ is disjoint. This transition occurs when the local maximum at $X^*=\sqrt{\frac{k_0}{3g}}$ reaches the value $2v_0$, i.e. when $2\left(\frac{k_0}{3g}\right)^{3/2}=v_0$, or $k_0=3g(v_0/2)^{2/3}$ (replacing $k_0$ by $k$ and taking $g=1$ and $v_0=1$ we recover $k_c$). For $N=2$ this transition is rather straightforward, but the fact that it persists at large $N$ is not a priori obvious. Note that for $N=2$ the particle density is always symmetric around the center of mass and thus $m_3$ is necessarily zero in this case. 
The symmetry breaking discussed in the text is thus a property which appears for larger values of $N$. 

In Fig.~\ref{fig:smallN} we show some examples of stationary densities $\rho_s(x)$ obtained numerically for $N=2,3,4,5$. Non-symmetric stationary states seem to become possible starting from $N=5$, which is consistent with the minimal value for the fraction of particles on the right $\alpha=1/3$ found in the text for $N\to+\infty$. Note that for $N=3$ the support of the density seems to be always connected.

These results show that the existence of multiple stationary states is not just an effect of large $N$ but is also present for small values of $N$ (e.g. for $N=5$ one may have either $2$ particles on the left and $3$ on the right or vice versa).

\section{Some more general interactions} \label{app:realistic}

\subsection{Numerical study of a case where the interaction force vanishes at infinity}

\begin{figure}
    \centering
    \includegraphics[width=0.48\linewidth]{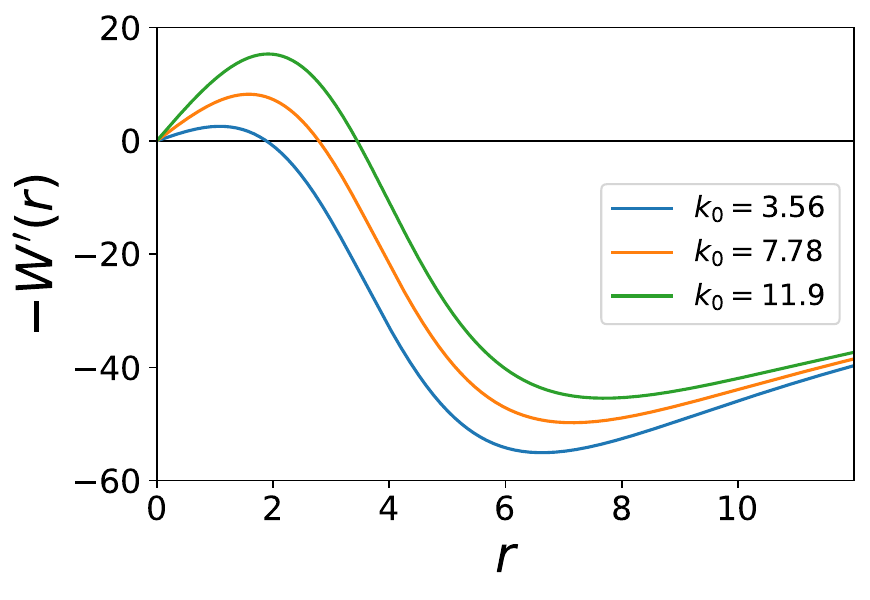}
    \includegraphics[width=0.45\linewidth]{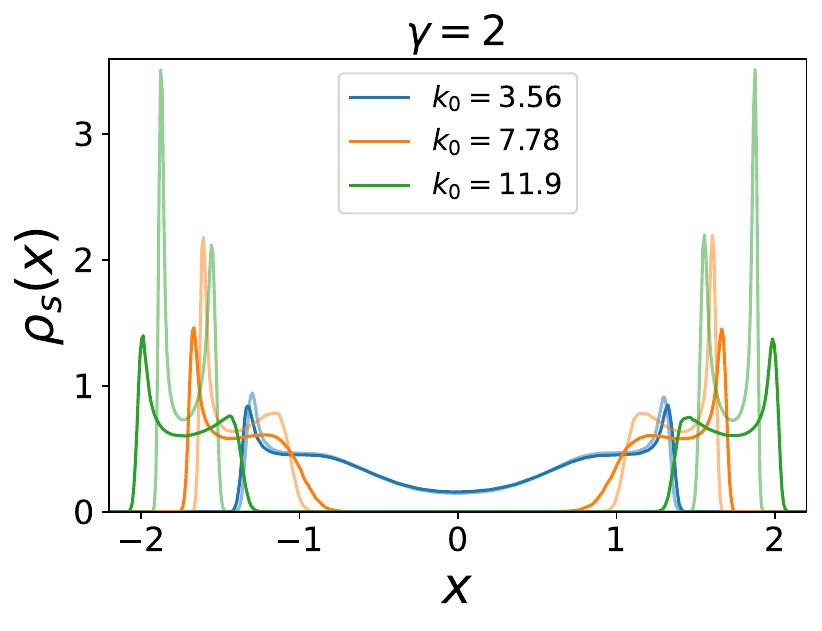}
    \caption{Left: Plot of the interaction force in \eqref{interaction_nodiv} for $g=1$, $a=0.002$ and different values of $k_0$ corresponding to the ones of Fig.~\ref{fig_density_symmetric}. Right: Plots of the stationary density $\rho_s(x)$ obtained from numerical simulations for $N=100$, $v_0=1$ and $\gamma=2$. The darker curves correspond to the force plotted on the left, while the lighter curves are the same as on the left panel of Fig.~\ref{fig_density_symmetric}, i.e. they correspond to the case $a=0$ with the same values of $k_0$. For small values of $k_0$ (blue), the two curves are very close to each other. For larger values of $k_0$ the differences become more important, but the behavior remains qualitatively similar.}
    \label{fig:realistic}
\end{figure}

As discussed in the main text, the fact that the interaction force diverges with the distance between the particles is somewhat unrealistic. A relevant question is therefore to what extent can our results be applied to situations where the force has a similar structure (repulsive at short distance and attractive at large distance), but does not exhibit such a divergence. To answer this question, we considered an interaction force of the form
\be \label{interaction_nodiv}
-W'(r)=\frac{k_0 \, r - g \, r^3}{1+a \, r^n} \;,
\ee
with $a>0$ and $n$ an even integer larger or equal to $4$, and performed numerical simulations for $N=100$ particles, see Fig.~\ref{fig:realistic}. The results shown here are for $n=4$, but other values of $n$ give similar results. We find that, as long as the absolute value of the minimum of the force is much larger than the value of the maximum, 
the shape of the density in the stationary state is qualitatively similar to what we found for $a=0$. In particular, we still observe a transition between a connected and a disconnected support when varying the parameter $k_0$.

\subsection{Limit $\gamma \to 0$}

The self-consistent equation for $\rho_s(x)$ can be analyzed analytically in the
limit $\gamma \to 0^+$ for a large class of interaction potentials $W(x)$. 
Indeed in that limit $\rho_s(x)$ becomes a sum of delta peaks,
as was noted in Sections \ref{sec:bistability} and \ref{subsec:gamma0asym},
and it is easy to evaluate the effective force 
$\tilde F(x)$. As in the rest of the paper, we assume that $W'(0)=0$ since 
the self-consistent equation was derived under this assumption.

\begin{figure}
    \centering
    \includegraphics[width=0.48\linewidth]{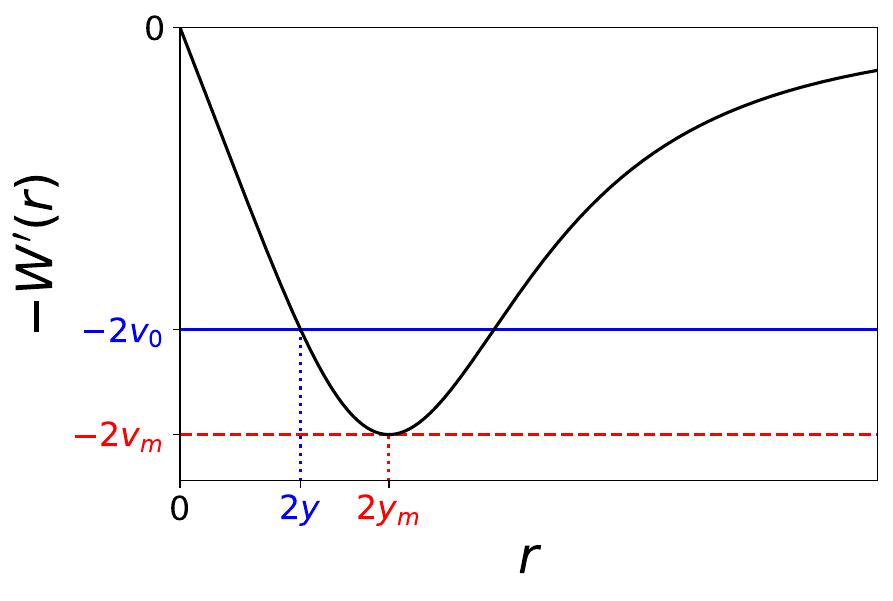}
    \includegraphics[width=0.46\linewidth]{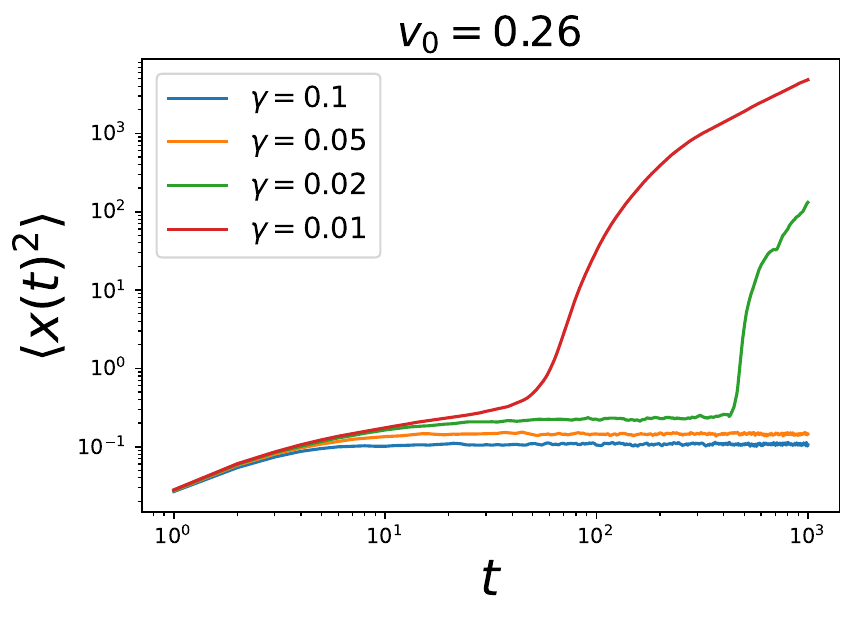}
    \caption{Example of a purely attractive force $-W'(r)$ which vanishes at $r=0$ and at large separation $r\to+\infty$. For a given value of $v_0$, the support of the stationary density of the bound state $[-y,y]$ is given by the smallest positive root of \eqref{eq_y_attractive}. The maximum value of $v_0$ for which such a root exist, $v_m$, is given by \eqref{eq_vm_attractive}. For $v_0>v_m$, there is a priori no bound state.
    Right: Second moment $\langle x(t)^2\rangle$ as a function of time (in log-log scale) for the interaction potential $W(x)= \frac{1}{2} \log(1+ x^2)$ with $v_0=0.26$ (just above $v_m=1/4$), for $N=100$ and for different values of $\gamma$. After a transient state which lasts longer as $\gamma$ increases, the particles seem to escape towards infinity.}
    \label{fig:attractive_ym}
\end{figure}

Let us consider first a purely attractive interaction force $- W'(x)$,
which vanishes at infinity, as depicted in Fig.~\ref{fig:attractive_ym}. Assume that there is
a bound state solution with a single support $[-y,y]$, $y>0$. Then as $\gamma \to 0^+$, the total density reads
\be
\rho_s(x) = \frac{1}{2} \delta(x-y) + \frac{1}{2} \delta(x+y) \;.
\ee
Hence, the total force acting on a particle at $x$ is
\be
\tilde F(x) = - \int dx' W'(x-x') \rho_s(x') = - \frac{1}{2} (W'(x-y) + W'(x+y)) =: \hat F(x,y) \;.
\ee
By definition, $y$ corresponds to the smallest positive root of the equation
\be \label{eq_y_attractive}
-v_0 = \tilde F(y)= \hat F(y,y) = - \frac{1}{2} W'(2 y) \;.
\ee 
This root exists and is unique for $v_0 < v_m$ defined by
\be \label{eq_vm_attractive}
v_m = \frac{1}{2} W'(2 y_m) \;,
\ee 
where $y_m$ is the first positive minimum of $- \frac{1}{2} W'(2 y)$,
i.e. the smallest positive root of $W''(2 y_m)=0$ (see Fig.~\ref{fig:attractive_ym}). 

For $v_0 > v_m$ there is no single support solution to the self-consistent equation,
and it is natural to conclude that there is no bound state.

{\it Example}. One example is $W(x)= \frac{1}{2} \log(1+ x^2)$, $-W'(x)=- x/(1+x^2)$. 
One finds $y= \frac{1}{8 v_0} (1- \sqrt{1- 16 v_0^2}\,)$ which converges
to $y \to y_m=1/2$ when $v_0 \to v_m = 1/4$. We have confirmed this in
a numerical simulation with $N=100$ particles and various small values of $\gamma$.
For $v_0>v_m$, we observe a transient regime, after which the particles escape to infinity (see the right panel of Fig.~\ref{fig:attractive_ym}). The duration of the transient state does not seem to depend on $N$,
but increases with $\gamma$. This can be understood as for smaller $\gamma$ the motion is more persistent which helps the particle to escape.
\\

Let us now consider the case where the potential is repulsive at short distance and 
attractive at large distance. One example is 
\be 
- W'(x) = \frac{k_0 x - g x^3}{1+ a  x^4} \;.
\ee 
{\it Solutions with a single support:} Assuming again a single support $[-y,y]$, $y>0$, the self-consistent equation for $y$ \eqref{eq_y_attractive} reads 
\be 
\tilde F(y)=  \hat F(y,y) = \frac{y}{1 + 16 a y^4} (k_0-4 g y^2) = - v_0 \;.
\ee 
This is possible for $y> \frac{1}{2}\sqrt{k_0/g}$. For $a=0$ one finds that 
$y=y$ is the largest root of $y (4 g y^2 - k_0)= v_0$, which coincides with \eqref{eq_y1_gamma0} after setting $g=1$ and $v_0=1$ as in the main text. 
For $a>0$ there is a maximum possible value $v_m$ for $v_0$ which is
\be 
v_m = \max_{y>0}  \frac{y}{1 + 16 a y^4} (4 g y^2-k_0) \;,
\ee 
beyond which there is a priori no bound state. It is easy to see that in the limit of small $a$
one has $y \sim 1/a$ and $v_m \sim 1/a^{3/4}$.

{\it Solutions with disjoint support:} For small $v_0$ we expect a disjoint support.
Assuming a disconnected (but still symmetric) support $[-y,-y']\cup[y',y]$, with $0<y'<y$, the density for $\gamma\to 0^+$ now reads
\be 
\rho_s(x) = \frac{1}{4} \delta(x-y) + \frac{1}{4} \delta(x-y') + \frac{1}{4} \delta(x+y') + \frac{1}{4} \delta(x+y') \;,
\ee
leading to
\be
\tilde F(x) = - \int dx' W'(x-x') \rho_s(x') =
- \frac{1}{4} (W'(x-y) + W'(x-y') +  W'(x+y) + W'(x+y') )  \;,
\ee
and the self-consistency equation \eqref{eq_y_attractive} becomes
\be 
\tilde F(y) = - v_0 \quad , \quad \tilde F(y') = v_0 \;.
\ee 
This gives
\bea  
&& - \frac{1}{4} (W'(y-y') +  W'(2 y) + W'(y+y') ) = - v_0 \;, \\
&& - \frac{1}{4} (W'(y'-y) +  W'(2 y') + W'(y+y') ) = v_0 \;.
\eea  

In the case $- W'(x)= k_0 x - x^3$ one finds, fixing again $g=1$ $v_0=1$ as in the main text,
\bea  
&& k_0 y-\frac{5  y^3}{2}-\frac{3 y (y')^2}{2}+1 = 0 \;,\\
&& k_0 y' -\frac{3 y^2 y'}{2}-\frac{5  (y')^3}{2}-1 = 0 \;,
\eea
which coincides with \eqref{eq_y1y3_gamma0_1}-\eqref{eq_y1y3_gamma0_2} with $y=y_1$ and $y'=-y_3$.
For $k_0$ large enough Mathematica finds some acceptable (i.e. real positive) roots. 
For instance for $k_0=9$, and $a=0$ one finds the only two possible acceptable roots
$(y,y')=( 1.70866,  1.23463)$ and $(y,y')= (1.93528, 0.319871)$.
Increasing $a$ and fixing $k_0=9$ these roots move as follows, for
$a=0.0002$ and $n=4$ one finds 
 $(y,y')=( 1.94692,  0.33173)$, and for 
$a=0.002$ one finds $(y,y')=( 2.03421, 0.522395)$.

To conclude we confirm that for small $a$ when $v_0$ increases there
is thus a bound state with a disjoint support, then a bound state
with a single support and finally, for $v>v_m$, no bound state. 

{\blue
\section{Values of $k_0$ for Fig.~\ref{fig_density_symmetric}} \label{app:table}

Table \ref{tab:k0k} gives the correspondence between the values of $k$ used for the analytical predictions and the values of $k_0$ used for the numerical simulations in Fig.~\ref{fig_density_symmetric}, as given by \eqref{def_k}.

\begin{table}[h!]
    \centering
    \begin{tabular}{|c|c|c|c|c|c|c|c|}
    \hline
        $k$ & -6.5 & -3 & -1 & 1 & 2 & 3 & 4 \\
        \hline
        $\gamma=2$ & & -2.862008 & -0.556904 & 3.561302 & 7.783972 & 11.895498 & \\ 
        \hline
        $\gamma=6$ & -6.475127 & & -0.803101 & 3.703725 & 7.900638 & 11.943203 & 15.962077 \\ 
        \hline
    \end{tabular}
    \caption{Correspondence between the values of $k$ used for the analytical predictions and the values of $k_0$ used for the numerical predictions on the two plots of Fig.~\ref{fig_density_symmetric}, computed using the relation \eqref{def_k}.}
    \label{tab:k0k}
\end{table}
}

\end{document}